\documentclass[12pt,a4paper]{article}
\usepackage[utf8]{inputenc}
\usepackage{amsmath}
\usepackage{amsfonts}
\usepackage{amssymb}
\usepackage{graphicx,caption}
\usepackage{mathtools,color}
\usepackage[margin=60pt]{geometry}
\newcommand {\be} {\begin{equation}} 
\newcommand {\ee} {\end{equation}} 
\newcommand {\Be} {\begin{eqnarray*}}
\newcommand {\Ee} {\end{eqnarray*}}
\newcommand {\bey}{\begin{eqnarray}} 
\newcommand {\eey}{\end{eqnarray}} 

\begin{document}

\title{\textbf{Thermalization of isolated harmonic networks under
conservative noise}}
\author{Stefano Lepri$^{1,*}$}
\date{%
    $^1$Consiglio Nazionale delle Ricerche, Istituto dei Sistemi Complessi\\
    via Madonna del piano 10, I-50019 Sesto fiorentino, Italy\\
    $^*$E-mail: stefano.lepri@isc.cnr.it\\[2ex]%
    \today
}

\maketitle
\abstract{We study a scalar harmonic network with pair interactions
and a binary collision rule, exchanging the momenta of a  
randomly-chosen couple of sites. We consider the case of the isolated
network where the total energy is conserved.   In the first part,
we recast the  dynamics as a stochastic map in normal modes (or action-angle)
coordinates and provide a geometric interpretation of it. 
We formulate the problem for generic networks but, for completeness, 
also reconsider the translation-invariant lattices. 
In the second part, we examine the kinetic limit and its range of validity.
A general form of the linear collision operator 
in terms of eigenstates of the network is given. This defines an 
\textit{action network}, whose connectivity gives information 
on the out-of-equilibrium dynamics. We present 
a few examples (ordered and disordered chains and elastic networks)
where the topology of connections 
in action spaces can be determined in a neat way. 
As an application, we consider the classic problem of relaxation 
to equipartition from the point of view of the dynamics of linear actions. We
compare the results based on the spectrum of the collision 
operator with numerical simulation, performed with a novel
scheme based on direct solution of the equations of motion 
in normal modes coordinates.}

\section{Introduction}

Nonequilibrium processes in many particle systems can be cathegorized 
in two wide classes: transport induced by external 
forces (either mechanical and thermodynamical)  and relaxation to equilibrium. 
In the first class,  we encounter steady-state transport,  or even time-dependent
states like it happens e.g.  in surface growth,  pattern formation, turbulent flows etc.  
The second class concerns problems like 
thermalization and approach to equipartition,  coarsening,  relaxation after a 
quench and so on.  Both themes continue to receive a large attention and 
the goal is to categorize them in broad classes characterized by 
some form of universality. 

The issue of thermalization has a long-standing history,  starting with 
the classic Fermi-Pasta-Ulam-Tsingou problem \cite{gallavotti2007fermi}
until the most recent developments, regarding the effect of 
integrability and quasi-integrability on the statistical 
behavior of classical systems
\cite{Benettin2013,DeRoeck2013,Huveneers2017,Fu2019,goldfriend2019equilibration,baldovin2021statistical}.  

One common trait in the study of such a problem is the difficulty of 
treating genuine anharmonic systems which are usually tackled by 
molecular dynamics simulations.  
This is the case of large ensembles 
of coupled nonlinear oscillators driven out of equilibrium
\cite{DHARREV,Lepri2016,benenti2020anomalous}. Even in the weakly-nonlinear limit, 
perturbative methods 
like the KAM theory are notoriously of little use in thermodynamically-large 
systems.  In the kinetic regime,  approaches based on the phonon 
Boltzmann  or wave-kinetic equation are well developed \cite{spohn2006phonon}.
For one-dimensional chains this turned out to be useful to understand
anomalous transport  \cite{Pereverzev2003,Nickel07,Lukkarinen2008,lukkarinen2016kinetic} 
and (pre)thermalization caused by nonlinear resonances \cite{Onorato2015,huveneers2020prethermalization}

An alternative approach relies on 
stochastic modeling of the mesoscopic dynamics, where the toolbox 
of stochastic processes can be fruitfully used. Considerable
insights is thereby obtained on the nature of the non-equilibrium
states and  macroscopic 
transport laws \cite{KMP82,Derrida1998}. 

An intermediate viewpoint is to consider
a sort of hybrid dynamics, where a deterministic evolution is 
accompanied by stochastic interactions,  possibly preserving the basic conservation
laws.  In the simplest case, the deterministic dynamics is linear and solvable, while the random part can be 
seen as  is a microcanonical Monte-Carlo rule, ergodizing the dynamics. 
This is referred to as conservative noise dynamics: in its simplest versions
it entails random exchange of momenta between particle or a random 
reshuffling of a subset of particles.   
Such  schemes,  allows sometimes for exact solutions or, at least,  very efficient simulations. For fluid systems, 
this is the strategy of the Multi-Particle-Collision dynamics that can be fruitfully
employed in many diverse contexts \cite{Malevanets1999,kapral08,DiCintio2017}.  
For the class of oscillator chains, this class of random dynamical systems
allows for an exhaustive rigorous treatment 
\cite{BBO06,Basile08,basile2016thermal,bernardin2012harmonic}.
Indeed, large-scale hydrodynamics equations can be demonstrated 
and  phonon Boltzmann equation can be derived, yielding relatively
simple linear collision operators  \cite{basile2010energy,lukkarinen2016harmonic}.
Moreover,  most of nonequilibrium steady-state
properties  can be computed exactly \cite{Lepri2009,Lepri10,Delfini10,Kundu2019}
and were shown to reproduce many features of deterministic nonlinear lattices \cite{Lepri2011b}. The effect of conservative noise on nonlinear
oscillator chains has also been considered \cite{Basile08,Iacobucci2010,Bernardin2014,lepri2020too}.

Another ingredient is the effect of quenched
disorder, leading to subtle interplay between Anderson localization 
and chaotic diffusion that affects both thermalization  and transport
\cite{Pikovsky08,Kopidakis08,Skokos2009,Lepri2010,Basko2011}.  A further 
motivation to tackle this case is that 
classical
disordered chains with nonlinear interactions exhibit a regime 
analogous to that seen in quantum many-body-localized systems \cite{kumar2020transport}.
In addition to intrinsic heterogeneity of individual units, disorder  
can originate also from  topology and structure  of connections, as it
occurs for the dynamics of elastic networks (see e.g. 
ref.\cite{bouchaud1990anomalous} and references therein).
This type of structures have an intrinsic theoretical interest, and 
may serve, for instance, as toy models of macromolecules (e.g. proteins)
in their native state \cite{tirion1996large,juanico2007discrete}.
Another possible domain of application is the study of nanoscale
heat transfer trough networked structures \cite{Freitas2014,Xiong2018},
namely  
as toy models of devices composed of networks of nanowires and
nanotubes.

When dealing with weakly-interacting linear systems it is natural to refer to the 
unperturbed harmonic modes and their energies or actions. 
For regular,  homogeneous,  anharmonic lattices 
these  are the usual Fourier modes.  Their perturbed dynamics determines 
the evolution of relevant observables both far \cite{Benettin2011}
than close  \cite{Lepri98c,L00} to equilibrium.  For disordered
lattices one has instead to consider Anderson modes that are typically localized 
in space.

Generally, \textcolor{black}{when expressed in the familiar
action-angle variables}, a nonintegrable perturbation defines a network of 
interaction among
the unperturbed actions  \cite{Mithun2018,danieli2019dynamical}. 
It is reasonable to argue that the connectivity of such network
will affect relaxation and ergodic properties. In general, one 
can distinguish such networks depending on whether  the number
of groups of actions linked by the perturbation depends
intensively or extensively on the number of degrees of freedom $N$. 
These are termed short or long-range networks, respectively, 
depending on whether the coupling range  if roughly constant or increases proportionally 
to $N$  \cite{Mithun2018,danieli2019dynamical}. 
In the nonlinear case the coupling may involve three or more actions, 
resulting in complicated hyper-graph structure.
It is thus of interest to characterize the connectivity of 
such action network in the simplest setting, 
and this is one of the aims of the present paper. 

In the present work, we present a general class of dynamical systems with 
conservative-noise: an harmonic network with general pair interaction 
described by a coupling matrix and equipped with a binary collision rule, 
exchanging the momenta of a couple of randomly chosen pair of particles.  
The motivations are twofold. First, we would like to
understand how the conservative-noise dynamics translates in the 
space of collective coordinates (i.e. the eigenstates of the underlying 
harmonic network).  In other words, how do local interactions in space 
affects the normal modes. Second, we would like to gain some
insights on the connectivity and topology of the resulting 
action networks, in the simplest setting. 
We also focus on the problem of relaxation 
to equilibrium concentrating on the dynamics of linear actions,  
a subject that has not been considered so far for conservative noise
dynamics.

In the first part,
we recast the stochastic dynamics as a stochastic 
map in normal modes (or action-angle)
coordinates and provide
a geometric interpretation of it. We will formulate the problem on 
generic scalar networks but, for completeness, we will reconsider also 
the translation-invariant lattices. As a byproduct, we derive 
an efficient and novel scheme to integrate numerically the equations
of motion in the collective coordinates.

In the second part, we examine the kinetic limit and its range of validity.
We compute the linear collision operator in the normal mode basis and 
discuss a few examples where the topology of connections 
in action spaces can be discussed in a neat way.
As an application, we study the classic problem of relaxation to equipartition
\textit{a l\'a} Fermi-Pasta-Ulam-Tsingou,  starting from an initial
condition with a few actions excited. We compare the exact dynamics
with the kinetic results.  A discussion and summary is given in the last
Section.

\section{Harmonic network with conservative noise}
       
Let us consider the following quadratic Hamiltonian,       
\begin{equation} 
\label{ham_all2all}
H=\sum_{i=1}^{N} \frac{p_{i}^{2}}{2m_i}+\frac{1}{2}\sum_{i,j=1}^N \Phi_{ij}q_{i}q_{j},
\end{equation}
where the matrix $\Phi$ is semi-positive definite and symmetric so that 
its eigenvalues are real and non-negative. The equation of motion of the isolated 
network are
\be
\ddot q_i = -\sum_j \Phi_{ij}q_{j}\quad .
\ee
In order to have momentum conservation it must be $\sum_i\Phi_{ij}=0$.  For 
simplicity we will deal henceforth with the equal-mass case,  and set $m_i=1$.

We now introduce the following stochastic process \cite{BBO06}. Suppose we start 
with the system at time $t$. At a later time $t+\tau$ there occur
a "collision" event defined as follows. A couple of particles $(m,n)$, $m\neq n$
is randomly selected according to the joint probability 
$W_{n,m}$ and their momenta are exchanged, $(p_n,p_{m})\to (p_{m},p_n)$. 
Clearly, this move conserves energy and the total momentum. 
Physically, it can be interpreted as a perfectly elastic collision 
as it would occur for a infinite square-well pair potential. 
In this general formulation,  the model encompasses several different setups.
It includes the standard case of regular Euclidean lattices when 
$\Phi$ is the familiar nearest-neighbor Laplacian matrix. 
The $W_{n,m}$ can be assigned to 
include some form of non-local interaction across the lattice. 
The intervals between subsequent collision times $\tau$ are also 
taken as random variables with some preassigned distribution 
with given, finite,  average $\langle \tau \rangle$. A natural 
choice would be, for instance, the exponential 
distribution $\exp (-\tau/\langle \tau \rangle)/\langle \tau \rangle$.

The normal modes' coordinates of the network have 
eigenfrequencies $\omega_\nu$ and are defined by the transformation 
\be
Q_\nu=\sum_{l=1}^N q_l \,\chi_l^{\nu}, \quad  P_\nu=\sum_{l=1}^N p_l \,\chi_l^{\nu}.
\label{modi}
\ee
The $\chi_l^{\nu}$, $\nu=1,\ldots,N$, are orthornormal and can be taken to be real
for the time being.
Once the Hamiltonian is expressed in these new canonical variables, the 
deterministic part of the equations of 
motion become
\bey
&&\dot Q_\nu \;=\; \frac{\partial H}{\partial P_\nu}=P_\nu , \\
&&\dot P_\nu \;=\; -\frac{\partial H}{\partial Q_\nu} =
-\omega_\nu^2 Q_\nu .
\label{newton}
\eey
Let us now determine the dynamics associated to  the random part. At each 
collision there is change in 
normal-mode momenta $\Delta P_\nu$  given by 
\[
\Delta P_\nu =  (p_{m}-p_n) (\chi_n^{\nu}-\chi_m^{\nu})
\]
while the $Q$ are unchanged. Using the inverse of (\ref{modi})
\[
p_n= \sum_{\nu} P_\nu  \chi_n^{\nu}, \quad
p_{m}-p_n  =  \sum_{\nu} P_\nu  (\chi_m^{\nu} -\chi_n^{\nu})
\]
we obtain
\[
\Delta P_\nu =   (\chi_n^{\nu} -\chi_m^{\nu})\sum_{\mu}(\chi_m^{\mu} -\chi_n^{\mu}) P_\mu .
\]
In the case in which $\Phi$ conserve the total momentum, 
momentum and stretch are automatically conserved since $\Delta P_0=0$.
So one can work with a $N-1$ variables. 
Altogether, in column vector notation, the collision rule
can be written as
\begin{gather}
P'=P+\Delta P=(1-2VV^T)P,  \\
 V_\nu^{(n,m)} \equiv \frac{\chi_n^{\nu} -\chi_m^{\nu}}{\sqrt{2}},
\label{DP} 
\end{gather}
where the prime denotes the value after the collision and $V^T$ the transpose.
Note that $V=V^{(n,m)}$ is a random vector as the indexes $n,m$ are chosen
at random with the prescribed rule (unless needed, we do not 
write explicitly the dependence on the site indexes 
henceforth). It can be checked that the vector $V$ is a unit vector, 
$V^TV=  1 $
Thus, equation (\ref{DP}) has an interesting geometrical interpretation. Indeed,  
$(1-2VV^T)$ it is a Householder matrix that describes a reflection about an 
hyperplane containing the origin. The hyperplane is orthogonal to the vector $V$.  
So the dynamics can be seen as a sequence  of reflections of the vector $P$
around a random hyper-plane in phase space,  a transformation that 
conserves the vector length $|P|^2=P^TP$.
The matrix properties are well known:
\begin{itemize}
\item it is idempotent $(1-2VV^T)^2=1$; 
\item it is self adjoint $(1-2VV^T)^+=(1-2VV^T)$; 
\item it has has  one 
eigenvalue equal to $-1$ and the remaining $N-1$ equal to $1$. 
 \item also $\Delta P = -2VV^TP = -2(P^TV) V$ where $R=P^TV=|P|\cos\varphi$ is the 
 projection of $P$ on $V$ so 
 $$
 P'=P -2RV  
 $$
 Since $|\Delta P|/|P|$ is of order one, the change in each component
 is of order $1/\sqrt{N}$. 
\end{itemize}


\section{Equations for the collective variables }

Let us now consider the dynamics in the usual action variables
to understand how the above transformations affects their dynamics.
Introducing 
\be
A=i(2\Omega)^{1/2}Q+(2\Omega)^{-1/2}P,
\label{Adefi}
\ee
with $\Omega_{\nu\nu'}=\omega_\nu \delta_{\nu\nu' }$
(note that $A_\nu$ are $2N$ independent complex variables).
The inverse formulae (recall that $P,Q$ are assumed to be real): 
\[
P=\frac12 (2\Omega)^{1/2}\left(A+ A^*\right);\quad 
Q=\frac{1}{2i} (2\Omega)^{-1/2}\left(A- A^*\right),
\]
with Hamiltonian transforming to 
$H=A^+\Omega A$. Substituting into (\ref{DP}) we obtain the collision
map:
\begin{equation}
A'= A -  M(A+A^*),
\label{Aprime}
\end{equation}
where the matrix $M$ is defined as
\[
M\equiv \Omega^{-1/2}VV^T\Omega^{1/2};\qquad M_{\nu,\nu'}= \sqrt{\frac{\omega_ {\nu'}}{\omega_\nu}} \,V_\nu V_{\nu'}
\]
($V$ is real). The  matrix $M$ is real, not symmetric since 
from the definition $M^T=\Omega M\Omega^{-1}$ and is idempotent 
$M^2=1$
expressing the fact that applying twice the same transformation 
does not change the state. 

For the collective variables, the free evolution  
in the time interval $(t,t+\tau)$ between subsequent collisions 
is diagonal
\begin{equation}
A(t+\tau) = e^{i\Omega \tau}A(t), \qquad
A^{*}(t+\tau) = e^{-i\Omega \tau} A^*(t).
\label{free}
\end{equation}
Combining the two processes (\ref{Aprime}) and (\ref{free}), 
we obtain an exact map from $t^+$ to
$(t+\tau)^+$
\begin{equation}
A \to (1-M)e^{i\Omega \tau}A-Me^{-i\Omega \tau}A^*.
\label{cmap}
\end{equation}
The evolution thus amounts to a sequence of multiplication 
of random matrices originating from the randomness of both the matrix 
$M$ and of the collision intervals $\tau$. This formulation is
particularly useful for implementing the numerical solution
as an event-driven dynamics, not requiring any approximate 
integration schemes as in the case of ordinary or stochastic 
differential equations (see  Appendix A).

As a further step, we seek for the equations for the action-angle 
variables $I, \theta$ defined by $A_\nu=\sqrt{I_\nu} e^{i\theta_\nu}$.
From equation (\ref{Aprime}) and recalling that $M$ is real, 
we obtain the collision map
\begin{eqnarray}
\label{Inu}
&&I_\nu' = I_\nu -4 \sqrt{\frac{I_\nu}{\omega_\nu}}V_\nu Z \cos \theta_\nu 
+ \frac{ 4V_\nu^2}{\omega_\nu} Z^2 \nonumber\\
&&=I_\nu \left[ \sin^2 \theta_\nu + \left(\cos \theta_\nu   - \frac{2V_\nu}{\sqrt{\omega_\nu I_\nu}} Z\right)^2 
\right]\nonumber\\
&& \sin \theta_\nu' = \sqrt{\frac{I_\nu}{I'_\nu}} \,\sin \theta_\nu\\
&&Z \equiv  \sum_\mu V_\mu \sqrt{\omega_{\mu}I_\mu} \cos \theta_\mu . \nonumber
\end{eqnarray}
\begin{itemize}
\item Those  equations are exact: the first one guarantees that 
the energy is conserved at each collision as $\sum_\nu \omega_\nu \Delta I_\nu =0$.
Note however that the total action $\sum_\nu I_\nu$ \textit{is not} a constant of motion.
\item The transformation made the collision map nonlinear, and 
$Z$ is global coupling among the eigenmodes: each collision entails 
a change of action and angles of all the modes.
\item  Finally, taking into account the free evolution, 
we can rewrite the dynamics as a mapping from $t^-$ to $(t+\tau)^-$
(i.e. just before two subsequent collisions)
\begin{gather*}
I \to  I + \Delta I(I,\theta)\\
\theta \to \theta + \omega\tau + \Delta \theta(I,\theta),
\end{gather*}
where $\Delta I=I'-I$ and $\Delta \theta$ are defined by the (\ref{Inu}) and $\omega$ 
is the vector of the eigenfrequencies $\omega_\nu$.  It is a 
a kind of random mapping or discrete Langevin equation, 
(but note that it is not symplectic). 
\end{itemize}
 
Before proceeding further, we note that 
the calculation can be performed also rewriting 
the equations (\ref{Aprime}) as a suitable stochastic equation.
The details are in Appendix B.

\section{Kinetic equation for the actions}

Let us now consider the kinetic limit, namely 
the case in which very many collisions occur on the 
time scale $T$, $T\gg \langle \tau \rangle$.
The limit in which $N \to \infty$, $\langle \tau \rangle \to 0$ keeping
\be
\gamma =\frac{1}{N\langle \tau \rangle}
\label{gamma}
\ee
finite, corresponds to choosing a finite collision probability per node of the 
network.  This is the choice mostly employed in the simulations \cite{Delfini10,Iacobucci2010,lepri2020too}. In this case, on the time-scale $T\sim 1/\gamma$ the number of collisions
becomes macroscopically large.
\textcolor{black}{Qualitatively, the phases
perform a random walk and are randomized on a faster time scale
with respect to the evolution of the actions. 
We thus argue that the angles $\theta_\nu$ are quickly randomized 
on this time scale and it is legitimate to average the first of 
(\ref{Inu}) over a uniform distribution of $\theta$, which is the
expected invariant measure.  Denoting by $\overline{I_\nu}$ 
such averaged actions we obtain
}
\begin{equation}
\overline{I_\nu'}=(1-2V_\nu^2)\overline{I_\nu} + 2\frac{V_\nu^2}{\omega_\nu}
\sum_{\nu'} \omega_ {\nu'}V_{\nu'}^2 \overline{I_{\nu'}}.
\label{Iprime}
\end{equation}
In other words, this last equation represents the evolution of an ensemble of trajectories
with a given distribution of initial angles, subject to the \textit{same 
sequence} of collisions.
Equivalently, in terms of the mode energies, $E_\nu=\omega_\nu \overline{I}_\nu$: 
\be
E_\nu'=(1-2V_\nu^2)E_\nu + 2V_\nu^2
\sum_{\nu'} V_{\nu'}^2 E_{\nu'} \equiv E_\nu + \sum_{\nu'} K_{\nu,\nu'}E_{\nu'}
\label{Eprime}
\ee 
which, taking into account the normalization of $V$, 
shows immediately that the sum of the $E_\nu$ is conserved also for the averaged
equation, 
and we have defined the angle-averaged collision matrix 
\begin{gather}
K_{\nu,\nu'}^{(n,m)}=-2V_\nu^2\,\delta_ {\nu,\nu'}+2 V_\nu^2  V_{\nu'}^2.
\label{colop}
\end{gather}
Each of the matrices $K$ has the following 
properties:
\begin{itemize}
\item It is a symmetric matrix and it is random as it depends on the couple of 
colliding particles $(n,m)$.  It does not seem to belong to any known ensemble 
of standard random matrices.
\item It does not depend on the eigenfrequencies and the 
distribution of collision times but only on the 
eigemodes' shape and on the collision probabilities $W_{n,m}$
\item It is seen immediately that $K$ admits a zero eigenvalue with
uniform eigenvector \textcolor{black}{$E_\nu=E_{eq}$ 
where $E_{eq}$ is the energy value of each mode at equipartition.}
\item $K$ is doubly-stochastic matrix, rows and columns sums to one.
\end{itemize}

We now would like to perform an average over realization of 
the collision process.
The formal solution from $t$ to $t+T$
is thus given in terms of a product of 
(uncorrelated) random matrices
\begin{equation}
E(t+T) = \left(\prod_{collisions (n,m)}(1+K^{(n,m)})\right)  \,E(t)
\label{formal}
\end{equation}
(with $1$ denoting here  the identity matrix) and the product
is extended to all the collision occurring in the 
time interval $t,t+T$.On the basis of the usual arguments based on Oseledec theorem, we 
thus expect that the associated Lyapunov exponents exist \cite{Pikovsky2016}.


If each application of the collision rule yields a small 
change in the vector $E$ (for instance if all elements of $K$ are 
small), one can approximate the product
in  (\ref{formal}) as 
\be
\prod_{collisions (n,m)}(1+K^{(n,m)}) \approx 1 + \sum_{collisions (n,m)}K^{(n,m)}.
\label{wdis}
\ee
\textcolor{black}{This approximation is akin to the 
well-known weak-disorder expansion, a method used to evaluate the 
product of random matrices for small disorder strengths \cite{Pikovsky2016}.
Its accuracy here will be afterwards checked numerically for the 
specific examples considered below. If we accept its validity for the 
time being we can,
as a further step, replace the sum in (\ref{wdis})} with the average $\overline{K}$ of the matrix over the random process. In other words, one considers the averaged operator
coarse-grained over a time scale which is longer than the typical collision
time:
\be
\overline{K}=\sum_{(n,m)}  W_{n,m} K^{(n,m)},\quad
R\equiv\frac{1}{\langle \tau \rangle}\overline{K} =\gamma N\overline{K}.  
\label{avmat}
\ee
\textcolor{black}{The rate $\gamma$ just sets the overall time-scale of the 
kinetic process.}
The (constant) matrix $\overline K$ inherits the properties of the
$K^{(n,m)}$.  Altogether, replacing the differences in (\ref{formal}) with 
a time derivative, 
the relaxation processes on such time scale is given by  
\be
\dot E_\nu= \sum_{\nu'}\left[R_{\nu,{\nu'}} E_{\nu'}
- R_{{\nu'},\nu}E_{\nu}\right]; \qquad
R_{\nu,{\nu'}}\equiv \frac{2}{\langle \tau \rangle}\overline{V_\nu ^2V_{\nu'}^2}.
\label{releq}
\ee
\textcolor{black}{This equation thus provides the seeked effective approximate evolution 
of mode energies on a time scale where the network has undergone 
a large number of collisions.}
Since energy is conserved, equation (\ref{releq}) can be seen as 
a Master Equation in action space and the elements of $R$ can be interpreted as 
transition probabilities from a state $E$ to $E'$. 
They are of the form of the well-known Fermi golden rule, involving
the squared amplitudes of the eigenmodes.
For a generic coupling of the network, such that there 
is no decoupling among different subsets they 
are all nonvanishing. Following the usual arguments of Markov
processes, the system is 
ergodic and approaches equipartition according to the 
properties of the "collision operator" defined by (\ref{releq}).
So the thermalization problem reduces to computing its 
$N$ eigenvalues $\mu_\nu$, \textcolor{black}{$\nu=1,\ldots,N$}, 
whose absolute values give the 
spectrum of relaxation rates. \textcolor{black}{The first eigenvalue $\mu_1=0$ 
and corresponds, as said,  to the steady state of energy 
equipartition. On physical grounds, $\mu_1$ must be non-degenerate 
since we expect the dynamics to be ergodic for a quadratic Hamiltonian
with such collision rules.}
All the others $\mu_\nu$ must be strictly negative \cite{schnakenberg1976network}.
In particular, it is of interest to look at the spectral 
gap $|\mu_2-\mu_1|=|\mu_2|$ that controls the relaxation 
at long times in a finite network.  The scaling with $N$ of 
$\mu_2$ (also called the Fiedler eigenvalue in the context of diffusion on
graphs)
can be recast in terms of the scaling of spectral density of the 
action network. \textcolor{black}{Let us denote with $\rho(\mu)$ the integrated
(cumulative) 
density of eigenvalues (i.e. the fraction of eigenvalues 
less than $\mu$).  If  
\be
\rho(\mu) \sim |\mu|^{d_\mu/2}, \qquad \mu \to 0 
\label{ds}
\ee
one may estimate $\mu_2$ for 
a finite network from the condition 
$\rho(\mu_2) =1/N$,
from which we obtain $|\mu_2| \sim N^{-d_\mu/2}$. 
Note that}, on general grounds,  $d_\mu$ is a distinct quantity from
the standard spectral dimension $d_s$ of 
the original network (\ref{ham_all2all}), which is defined
by the same relation as (\ref{ds}) for 
the integrated density of eigenfrequencies $\omega_\nu$ 
\cite{burioni1996universal,burioni2005random}.


Before proceeding further, let us comment on the validity of the weak-disorder approximations (\ref{wdis}).
It rests on the fact
that the change of the energy vector is somehow small at each collision. 
This is plausibly 
true in the case of 
extended modes, as it occur in the case of translation-invariant 
or weakly disordered lattices that will be treated below: each $\chi$ is 
of order $1/\sqrt{N}$ for normalization. From the very 
definition of the $V$, equation (\ref{DP}), we see that the matrix
elements of the relaxation matrix should be small for 
large $N$. In other words, each local collision (which 
yield a change $O(1)$ of the particle momenta) yield a $O(1/N)$
change of the mode energies. This expectation will be made 
precise in the standard case of the ordered chain (see 
Section \ref{sec:transl}).

On the other hand, this may not be true in general, e.g if the network admits 
localized modes. To understand this issue, let us consider 
the most trivial example in which $\Phi$ is diagonal
with frequencies $\omega_\nu = \sqrt{\Phi}$ and trivial 
eigenvectors $\chi_m^{\nu}=\delta_{m,\nu}$ (assuming that
they are labeled according to their spatial location). 
The matrix $K$ can be worked out: it is all zeros 
except for two diagonal elements and two elements in positions
$(n,m),(m,n)$, 
\be
K_{\mu,\nu}^{(n,m)} =  
\begin{cases*}
-\frac12  & if  $(\mu,\nu)=(m,m),(n,n)$  \\
\phantom{-}\frac12  & if $(\mu,\nu)=(m,n),(n,m) $\\
0 & otherwise.
\end{cases*} 
\label{indep}
\ee
Thus, the matrix is very sparse with
$O(1)$ entries, very different from the case of extended plane waves.
However the change involves only one mode, thus resulting in 
a relatively small change of the vector $E$ even in this case.
\footnote{Actually, this extreme case is basically 
the well-known KMP model whereby harmonic oscillators exchange 
energy stochastically \cite{KMP82}. But, it  may be also related
to deterministic models like the linear chain of oscillators 
with hard-core collisions, termed the ding-dong model \cite{Prosen92}
in its disordered version \cite{pikovsky2020scaling}.
It would correspond to the "stochastic approximation" of the dynamics
whereby the deterministic sequence of collisions is
replaced by a random one. } 

For a general $\Phi$ the eigenvectors are not known analytically
but can be easily computed numerically by exact diagonalization.
Then, one can compute the matrix form of the collision operator
by averaging over the chosen distribution $W_{n,m}$.  
The structure of the matrix will yield information
on the structure of action network.
An analytically treatable case is the one 
of ordinary lattices that we will discuss below.

\section{Translation-invariant lattice}
\label{sec:transl}

We now consider the case of a translation-invariant lattice,  focusing
on a one-dimensional chain with periodic boundary conditions. In this
case of $\Phi$ is a circulant matrix: this is 
a well-studied case both for short (nearest-neighbour) 
\cite{BBO06} than for long-range interactions \cite{tamaki2020energy}.
In both cases the eigenvectors
are the familiar lattice Fourier modes 
\[
\chi_l^\nu = \frac{1}{\sqrt{N}}e^{ik_\nu l}; \quad
V_ \nu \equiv \frac{ e^{ik_\nu n} -e^{ik_\nu m}}{\sqrt{2N}}
\]
where $k_\nu=\frac{2\pi \nu}{N}$ are the wavenumbers and
\be
Q_\nu={1\over\sqrt{N}}\sum_{l=1}^N q_l \, e^{ik_\nu l},\qquad
Q_{-\nu}=Q_\nu^*, \qquad \nu=-{N\over2}+1,\ldots, {N\over2}.
\label{modit}
\ee
(for simplicity of notation we use matrix indexes in the same range).  
A difference with the case above is that 
the vectors $V$ are complex. This requires some minor modifications
when expressing the equation of motion in phononic variables.
As explained in Appendix A, the collision map (\ref{cmap}) now reads
\begin{equation}
A \to (1-M)e^{i\Omega \tau}A-Me^{-i\Omega \tau}\tilde A^*,
\label{cmapt}
\end{equation}
where we introduced the shorthand notation $(\tilde A^*)_\nu=A_{-\nu}$.
This makes transparent how the collision involves scattering of phonons 
with opposite propagation directions.
The matrix $M$ is formally as before but is now complex-valued
\be
M\equiv \Omega^{-1/2}VV^+\Omega^{1/2};\qquad M_{\nu,\nu'}= \sqrt{\frac{\omega_ {\nu'}}{\omega_\nu}} \,V_\nu V_{\nu'}^*
\label{Mcomplex}
\ee
and it is non-Hermitean and idempotent. 

The collision map in action-angle variables is:
\begin{eqnarray}
&&I_\nu' = I_\nu -4 \sqrt{\frac{I_\nu}{\omega_\nu}} Re(V_\nu  e^{-i\theta_\nu}) Z 
+ \frac{ 4|V_\nu|^2}{\omega_\nu} Z^2=\nonumber\\
&& I_\nu \left|1-
 \frac{ 2V_\nu e^{-i\theta_\nu}}{\sqrt{I_\nu \omega_\nu}} Z\right| ^2\\
&& \sin \theta_\nu' = \sqrt{\frac{I_\nu}{I'_\nu}} \,\sin \theta_\nu -2 
\frac{Im(V_\nu)}{ \sqrt{ \omega_\nu I'_\nu}}Z\\
&&Z \equiv \frac12 \sum_\mu \sqrt{\omega_{\mu}I_\mu} (V_\mu e^{-i\theta_\nu} + V_\mu^* e^{i\theta_\nu})\nonumber
\label{Inuc}
\end{eqnarray}
(note that $Z$ is real). 

Accordingly, in the kinetic limit we can proceed with the averaging as above. The 
resulting kinetic equations 
are equal to  (\ref{Iprime},\ref{Eprime}) and (\ref{colop}) with the $V_\nu^2$ replaced by
\begin{equation}
|V_\nu|^2= \frac{1}{2N}|1-e^{ik_\nu (m-n)}|^2. 
\label{Vchain}
\end{equation}

With respect to the general network, for linear chains there are some important simplifications.
\begin{enumerate}
\item In the standard case of nearest-neighbor collisions 
$m=n+1$ so that the above quantity is 
independent of $n,m$ 
$$
|V_\nu|^2= \frac{1}{2N}|1-e^{ik_\nu}|^2 = \frac{2}{N}\sin^2\frac{k_\nu}{2},
$$
and the matrix $K$ is constant. Note that 
this remains true even for any fixed-distance collision rule
$l= m-n$.  The collision operator is 
determined straightforwardly
from definitions (\ref{gamma}) and (\ref{avmat})
\begin{equation}
R_{\nu,\nu'}=\gamma N {K}_{\nu,\nu'}
=-4\gamma\sin^2\frac{k_\nu}{2} \delta _{\nu,\nu'}+ 
\frac{8\gamma}{N} \sin^2\frac{k_\nu}{2}\sin^2\frac{k_{\nu'}}{2}
\label{Rnn}
\end{equation}
Since the off-diagonal terms are small for large $N$, the 
eigenvalues of the matrix are well approximated by the diagonal
elements giving the approximation
\be
\mu_\nu \approx -4\gamma\sin^2\frac{\pi (\nu-1)}{N},
\label{mu} 
\ee
up to higher-order corrections in $1/N$ that may be computed perturbatively.  
Also,  the $\nu$th eigenvectors is localized on $\pm \nu$ with all other 
components being small of order $1/N$. 
Since the spectrum has a vanishing gap for large $N$, the relaxation rate of 
a generic non-equilibrium initial condition is expected to occur on  
of the slowest time scale $\mu_2\approx\gamma (\pi/N)^2$ at long times.

\item A more general model would consist to consider a collision 
probability of the form
\begin{equation}
W_{n,m} = \, w_{n-m} 
\label{wnm}
\end{equation} 
meaning that we choose an random particle $n$ with equal probability and 
a neighbor at distance $l=n-m$ with probability $w_l$ 
(with periodic boundary condition assumed), 
the standard case above being $w_l =\delta_{l,1}$.
In this case,  the matrix $K$ is no longer constant. 
The averaged collision operator is expressed as
\begin{equation}
R_{\nu,\nu'}=
\gamma N {\overline K}_{\nu,\nu'}=  
\gamma \sum_l w_l
\left[
-4\sin^2\frac{k_\nu l}{2} + 
\frac{8}{N} \sin^2\frac{k_\nu l}{2}\sin^2\frac{k_{\nu'}l}{2}
\right]
\label{Rwnm}
\end{equation}

which, as above, suggests the following approximation in terms of the 
diagonal elements
\be
\mu_\nu \approx  -{4\gamma} \sum_l w_l.
\sin^2\frac{k_\nu l}{2}.
\label{diag2}
\ee
We will discuss an example later on.
\end{enumerate}

\textcolor{black}{Before concluding this Section, we note that the choice of 
the $w_{l}$ is crucial to assess the dependence of the $\mu_\nu$ on 
the wavenumber $k_\nu$, as seen already at the level of the approximation (\ref{diag2}).
In particular, introducing correlations and/or long-ranged rules 
may yield non-standard relaxation and transport depending on the range
and correlation strengths. We will defer investigation of those issues 
to future works.}

\section{Examples}

In this section we illustrate the above in some specific examples and check
for the validity of the approximations done.

\subsection{Disordered chain}

For a first test, let us  consider the case of the 
harmonic chain with disorder in the pinning potential
\be 
H = \sum_{i=1}^N \left[{p_i\over 2}^2
+\frac12 (1+\sigma_i) q_i^2 +
\frac{1}{2}(q_{i+1}-q_{i})^2\right] 
\label{dishami}
\ee
with $\sigma_i$ being i.i.d. variables with 
uniform distributions in $[0,w]$, and $w$ gauges the disorder 
strength. Periodic boundary conditions are assumed.
As it is well known, the eigenstates are the exponentially-localized 
Anderson modes, whose localization length decreases with $w$ \cite{Matsuda70,Visscher71}. We consider the standard case of nearest-neighbor collisions. Several variants in this type of model have been considered earlier
\cite{bernardin2008thermal,dhar2011heat,bernardin2019hydrodynamic}.

\begin{figure}
\begin{center}
\includegraphics[scale=0.6]{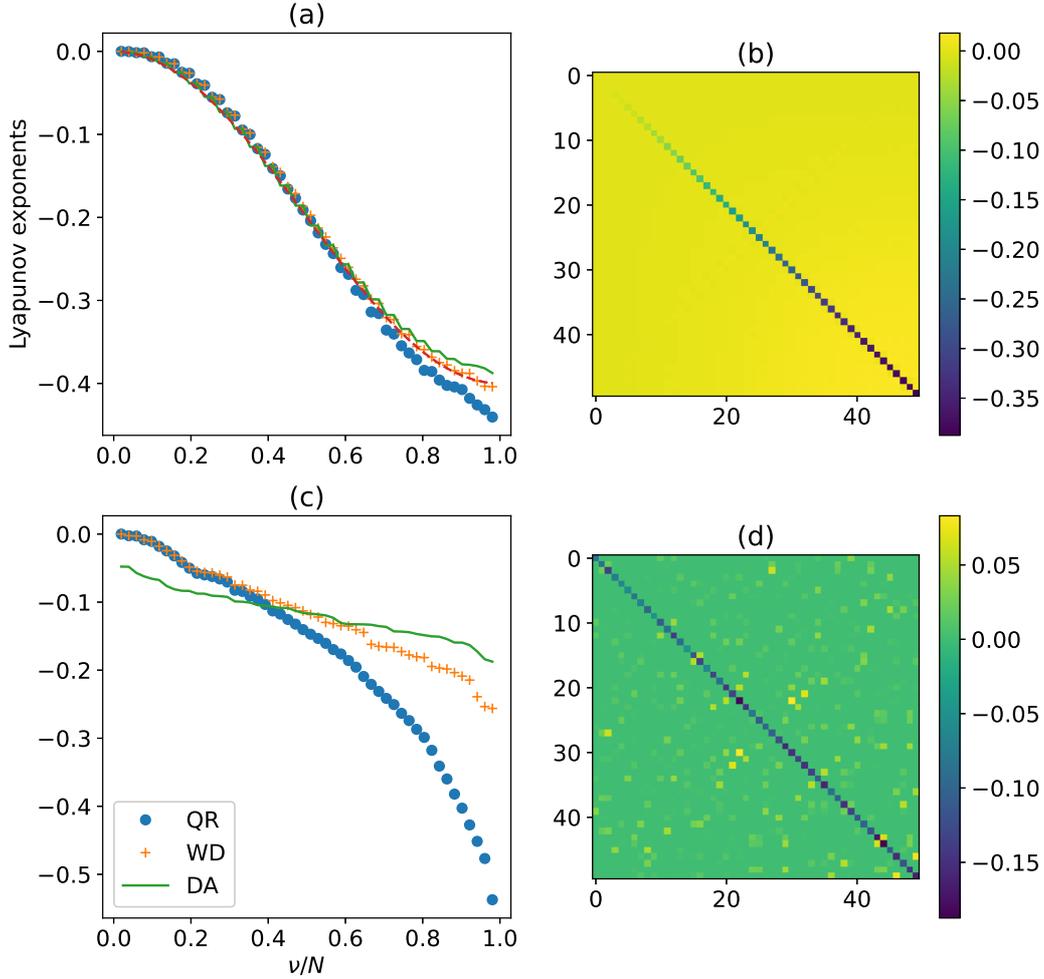} 
\end{center}
\caption{Disordered chain (\ref{dishami}), 
$N=50$, $\gamma=0.1$ with nearest neighbour collisions 
and disorder strengths $w=0.1$ (a,b) $w=10$ (c,d). 
Panels (a,c) report the  Lyapunov exponents of 
equation (\ref{releq}): Circles are the numerical
values computed via the QR algorithm, crosses are from the weak-disorder
expansion (eigenvalues of the average matrix, as in equation (\ref{wdis}))
The solid lines are the diagonal elements of the matrix itself.
The dashed in panel (a) line are the diagonal elements in the case 
of no disorder $w=0$, equation (\ref{mu}).
Panels (b,d): color plot of the averaged collision matrix
(\ref{avmat}).}
\label{fig:lyapdis}
\end{figure}

In figure \ref{fig:lyapdis} we illustrate the kinetic regimes
for two cases corresponding to relatively weak and strong 
disorder. \textcolor{black}{We first of all compute the 
relaxation spectrum as  
the Lyapunov exponents of the product or 
random matrices (\ref{formal}) via the usual QR algorithm
\cite{Pikovsky2016}. Then, we compare the result
with the eigenvalues of the averaged collision matrix
as obtained from the weak-disorder expansion (\ref{wdis}).} In the case of 
weak disorder, figure \ref{fig:lyapdis}(a) , the two methods give the 
same spectra, thus confirming the accuracy of the approximation.
Moreover, the eigenvalues are also very close to the diagonal 
elements which, in turn, are almost indistinguishable from the 
one of the ordered case (see below). This means that the collision operator
is almost diagonal 
with very small off-diagonal elements (see figure \ref{fig:lyapdis}b).

The situation is different in the strong disorder case. Here, 
only first 25\% of Lyapunov exponents coincide with the eigenvalues 
coincide and they are both very different from the 
diagonal elements (see figure \ref{fig:lyapdis}c). 
This is presumably due to the fact that the dynamics is 
comparatively less "mixing" and stronger deviations from the 
average occur on shorter times (those corresponding 
to more negative Lyapunov exponents).
Accordingly, the structure of the collision matrix 
is comparatively less homogeneous, with relatively larger 
off-diagonal elements (figure \ref{fig:lyapdis}d).

\begin{figure}
\begin{center}
\includegraphics[scale=0.8]{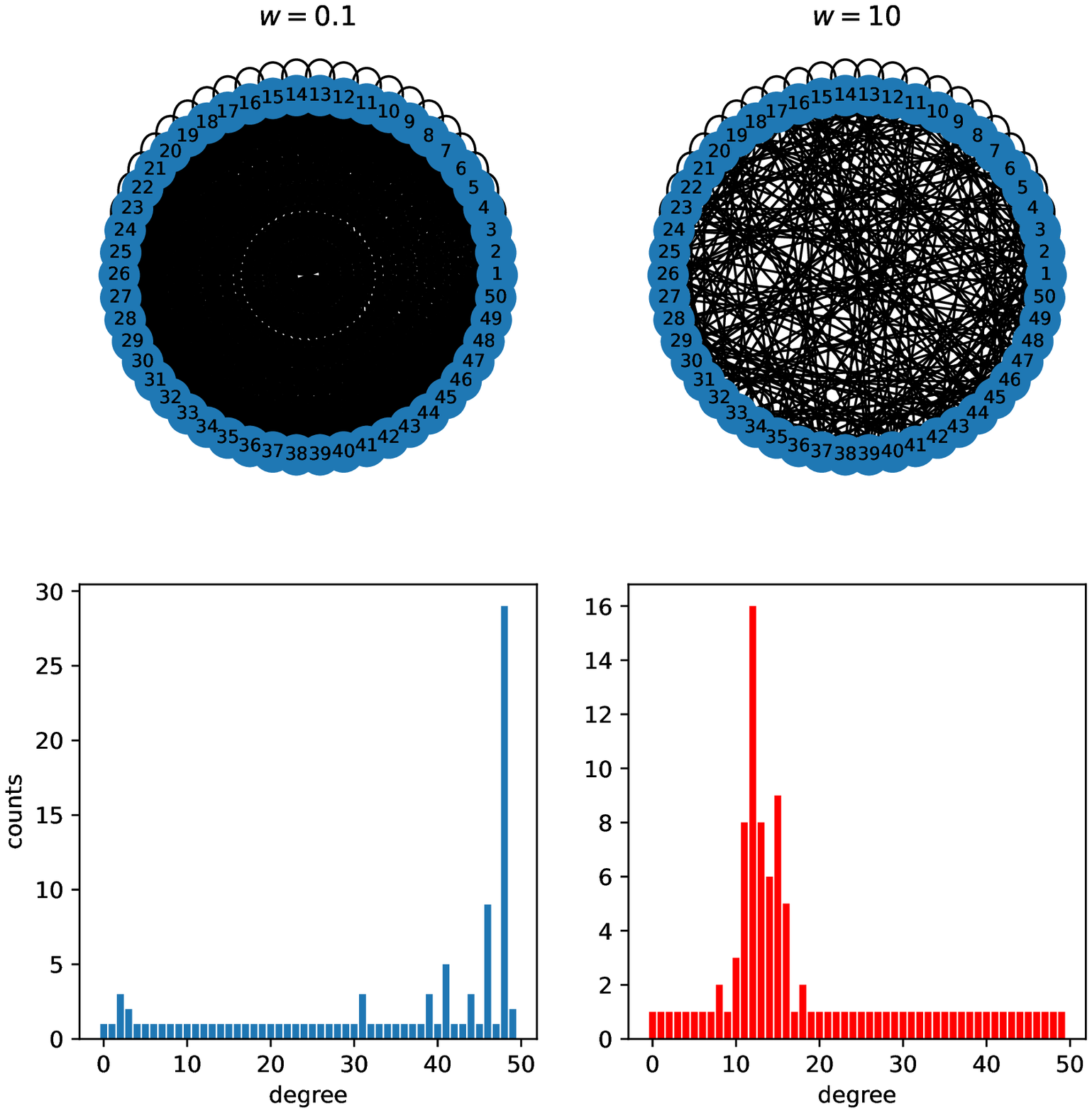} 
\end{center}
\caption{Disordered chain (\ref{dishami}), 
$N=50$, $\gamma=0.1$ with nearest neighbour collisions 
and disorder strengths $w=0.1$ (a,b) $w=10$ (c,d). 
(a,c) Graphical representation of the action network: 
it corresponds to an adjacency matrix where two nodes $\nu,\nu'$
are consider to be connected if $|R_{\nu,\nu'}|>10^{-4}$. 
(b,d): degree distributions.}
\label{fig:anet}
\end{figure}

A further insight is obtained by looking at the connectivity
in action space. As said above, the transition rates 
are all-nonvanishing, but can be very small due to the 
exponential localization of the eigenvectors. 
For the sake of visualization, 
we draw a representation of action network  
as an adjacency matrix as follows. We consider two modes $\nu,\nu'$ to be 
connected if $|R_{\nu,\nu'}|$ is larger than some preassigned threshold
and not-connected otherwise.
As seen in figure \ref{fig:anet}
there is a clear qualitative difference between the two cases:
increasing disorder the connectivity \textcolor{black}{decreases} and passes from 
a complete-graph type to a more sparse structure.
Accordingly, the degree distributions (the histogram of the 
number of links of each node) is distinctly
different. For weak disorder, each action is connected
to $O(N)$ ones, while the average degree is finite in the 
strong disorder case.

\subsection{Mean-field chain}

To test also the case of translation-invariant lattice we consider, 
for the sake of an example,   a standard ordered  chain with nearest-neighbour
with an additional mean-field  interaction: 
\be 
H = \sum_{i=1}^N \left[{p_i\over 2}^2+\frac{1}{2}(q_{i+1}-q_{i})^2\right] 
+\frac{k}{2N}\sum_{ij}(q_{i}-q_{j})^{2}.
\label{hami}
\ee
This type of chains has been discussed as a model for Bose-Einstein 
condensate \cite{yan2003harmonic} and studied recently in a nonequilibrium setup \cite{defaveri2021heat,andreucci2021classical}.
The eigenfrequencies are 
\be
\omega_\nu^2 \;=\; 2k + 4  \sin^2\left({\pi \nu\over N}\right) \quad.
\label{barefreq}
\ee
for $\nu\neq 0$ and $\omega_0=0$. Apart from this zero mode,  the 
dispersion is formally the same as the one of a chain with harmonic pinning.

As a result of the calculation in Section
\ref{sec:transl}, the kinetic equations are independent 
on $k$ (this will be checked later on).  For the sake of testing
the validity of the kinetic approximation, we here will 
concentrate on case $k=0$, which is well-studied and sometimes referred to as the 
acoustic chain, since the dispersion of waves is linear for 
small wavenumbers. We will profit also to test the effect of different 
collision rules of type (\ref{wnm}).

\begin{figure}
\begin{center}
\includegraphics[scale=0.6]{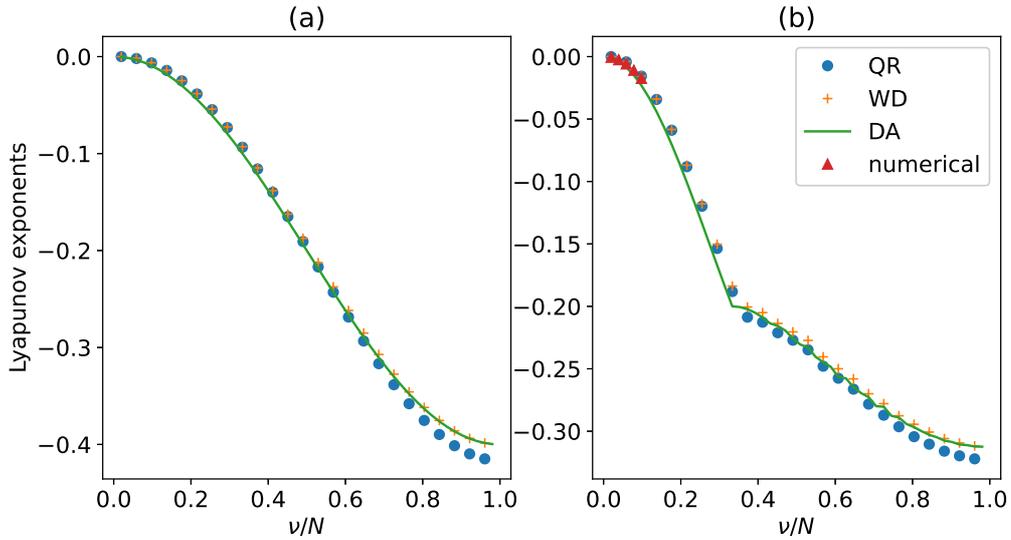} 
\end{center}
\caption{Lyapunov exponents of equation (\ref{releq}) 
for the case of the translation-invariant acoustic chain (\ref{hami}) with 
$k=0$, $N=50$, $\gamma=0.1$ and two types of for collision rule:
among (a) nearest-neighbor   
and (b) nearest-neighbor and next-to-nearest neighbor sites with 
probability $s=1/2$ each, equation (\ref{wnm}. Circles are the numerical
values of the Lyapunov exponents  computed via the QR algorithm, crosses are from the weak-disorder
expansion namely the eigenvalues $\mu_\nu$ of the averaged matrix, as in equation (\ref{wdis}).
Solid lines are the diagonal approximation given by
(\ref{mu}) and (\ref{diag2}). To take into account the double degeneracy, only 
half of the exponents is plotted.
Finally, the triangles are the relaxation rates measured 
from a simulation of a chain with the same parameters and $N=128$
as described in Section \ref{sec:equip}.}
\label{fig:lyaps}
\end{figure}

We first start with the relaxation rates obtained by the kinetic approximation.
As above,  we computed the spectrum numerically using the 
QR decomposition method \cite{Pikovsky2016} and compare it
with the weak-disorder expansion and diagonal approximation. 
In figure \ref{fig:lyaps} we report the 
relaxation rates computed with the different methods in two 
instances of (\ref{wnm}) ,
namely the standard nearest-neighbor exchange and the one  
with a dichotomic choice 
between both the nearest and next-to-nearest neighbor,
with probabilities $s,1-s$ respectively,
\be
w_l=s\delta_{l,1}+(1-s)\delta_{l,2}.
\label{wdico}
\ee
In the first case,  the matrix $R$ is given by (\ref{Rnn}) while 
in the second we need to compute
the Lyapunov spectra of the product of two constant matrices,
randomly chosen with probability $s$.
This correspond to the case of 
binary disorder in the random matrix language \cite{Pikovsky2016}. 

As illustrated in figure \ref{fig:lyaps} the three methods 
agree fairly well for both collision rules. We examined also 
other values of $k$ (data not reported) obtaining identical 
results as predicted.  It should also be noticed that the 
diagonal approximation does reproduce quite accurately the 
spectrum already for $N=50$. 

\subsection{Elastic network}

As an example of a more complicate topology, 
we consider a more general 
elastic network \cite{bouchaud1990anomalous,hastings2003random}
 \be 
H = \sum_{i=1}^N \left[{p_i\over 2}^2+\frac{1}{2}(q_{i+1}-q_{i})^2\right] 
+\frac{1}{2}\sum_{ij}C_{i,j}(q_{i}-q_{j})^{2},
\label{sparse}
\ee
where $C_{i,j}=1$ if the nodes are connected and zero otherwise.  
As a case study, we choose the well-known Newman-Watts-Strogatz network,
generated starting from a ring where each oscillator 
is coupled to its neighbor \cite{newman1999renormalization}. 
Each oscillator is then connected by to with 
randomly-chosen existing one with probability $p$. 
Increasing the probability $p$ increases the level of disorder and we may 
expect a change the 
normal mode structure from extended plane-waves to localized. 
The spectral properties have been studied in great detail in
\cite{monasson1999diffusion}.  The main feature is the 
existence of a pseudo-gap with very small eigenfrequencies
below a certain value.

\begin{figure}
\begin{center}
\includegraphics[scale=0.60]{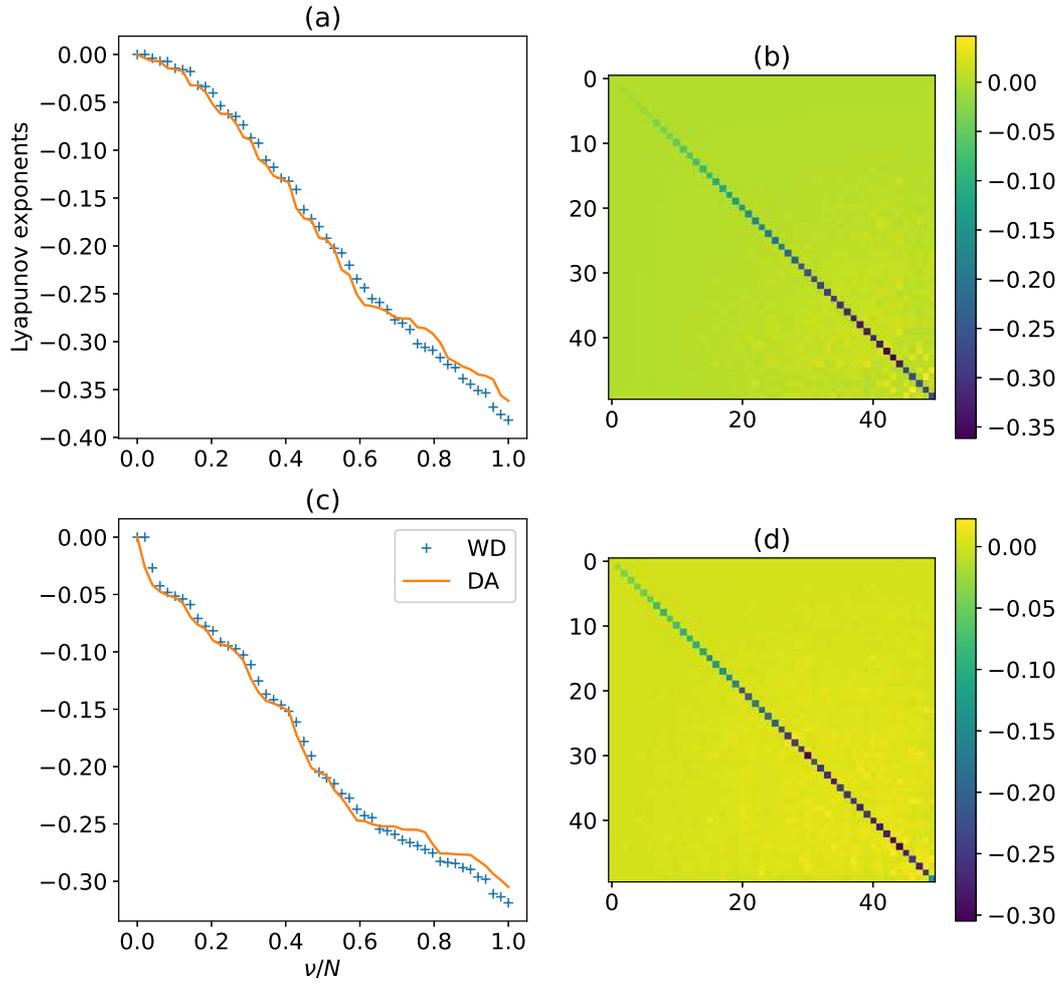}
\end{center}
\caption{Watts-Newman-Strogatz elastic network 
(\ref{sparse}) $N=50$, $\gamma=0.1$ with nearest-neighbour collisions 
and  $p=0.1$ (a,b) $p=0.8$ (c,d). 
Panels (a,c) report the  Lyapunov exponents of 
equation (\ref{releq}):  crosses are from the weak-disorder
expansion (eigenvalues of the average matrix, as in equation (\ref{wdis}))
The solid lines are the diagonal elements of the matrix itself.
Panels (b,d): color plot of the averaged collision matrix
(\ref{avmat}).} 
\label{fig:spar}
\end{figure}

In figure \ref{fig:spar} we we compare
for two cases corresponding to a single realization of $C_{i,j}$ 
and for relatively weak and strong 
disorder. We limit ourselves to compare  
the eigenvalues of the averaged collision matrix
from the weak-disorder expansion (\ref{wdis}) with the diagonal
elements. In both examples, the two are 
pretty close, especially the smallest ones. 
This means that the collision operator
is almost diagonal with very small off-diagonal elements,
similar to the case of the strongly disordered chain.  

\begin{figure}
\begin{center}
\includegraphics[scale=0.8]{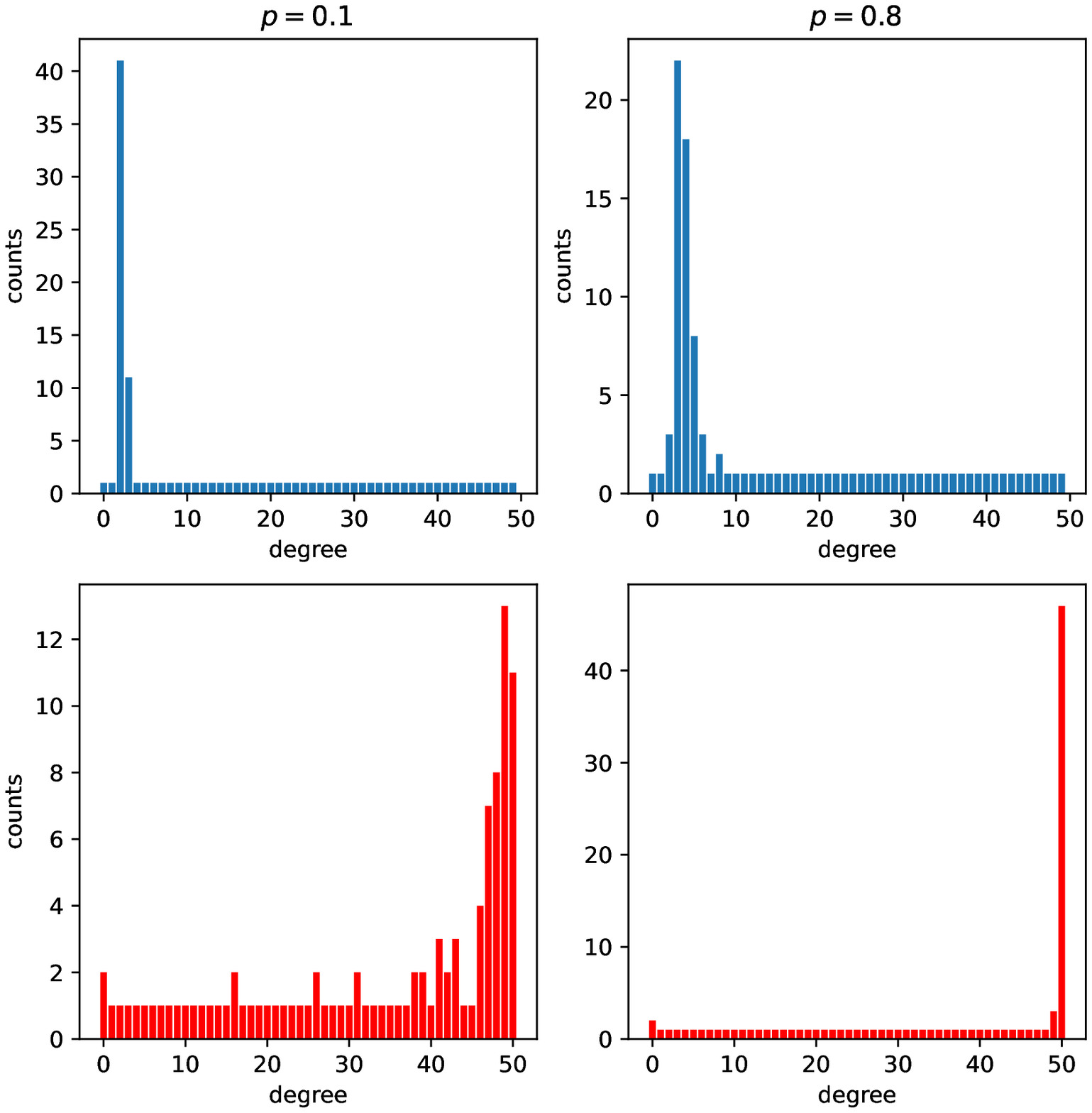} 
\end{center}
\caption{Watts-Newman-Strogatz elastic network (\ref{dishami}), 
$N=50$, $\gamma=0.1$ with nearest-neighbour collisions 
and  $p=0.1$ (a,b) $p=0.8$ (c,d). 
Comparison of degree distributions in the real network
(a,b) and in the the action network (c,d): 
it corresponds to an adjacency matrix where two nodes $\nu,\nu'$
are considered to be connected if $|R_{\nu,\nu'}|>10^{-4}$. 
.}
\label{fig:anet2}
\end{figure}

In figure \ref{fig:anet2} we compare the degree distribution
in the real network with the action network (as defined 
above) for two values
of $p$. As it is well known, the degree distribution of the is 
peaked around the average value \cite{newman1999renormalization}. 
On the contrary, 
the action network, constructed as described above, 
is fully connected, with practically all-to-all couplings.

\section{Thermalization}
\label{sec:equip}

We perform a thermalization simulation in the spirit of the 
Fermi-Pasta-Ulam-Tsingou famous numerical experiments.
We initialize the chain by exciting a packet of $N_0$ 
normal modes to have
a prescribed energy, namely we set
\be
A_\nu(0) = \sqrt{\frac{E_0}{2N_0\omega_\nu}} e^{i\alpha_\nu}
\label{Aini}
\ee
for $0<\nu\le N_0$ ($|\nu|\le N_0/2,\nu \ne 0$ in the case of translation-invariant chain) where $\alpha_\nu$ being i.i.d.random phases uniformly 
distributed in $[0,2\pi]$. The remaining modes are initialized as $A_\nu(0)=0$.
The dynamics is evolved iterating the 
dynamics as prescribed, with random Poissonian collision 
times whose average is given by (\ref{gamma}).
 
The evolution of the mode energies $E_\nu$ towards 
equipartition is monitored and compared 
with the predictions. To improve the statistical accuracy 
a time-average over a fixed number of collision (typically $10^3$) 
is performed, along with an average over different initial
conditions, i.e. different realization of $\alpha_\nu$ in (\ref{Aini}).
At long enough times, the average mode energies converge to 
the equipartition value as they should.

\subsection{Disordered chain}

\begin{figure}
\begin{center}
\includegraphics[scale=0.6]{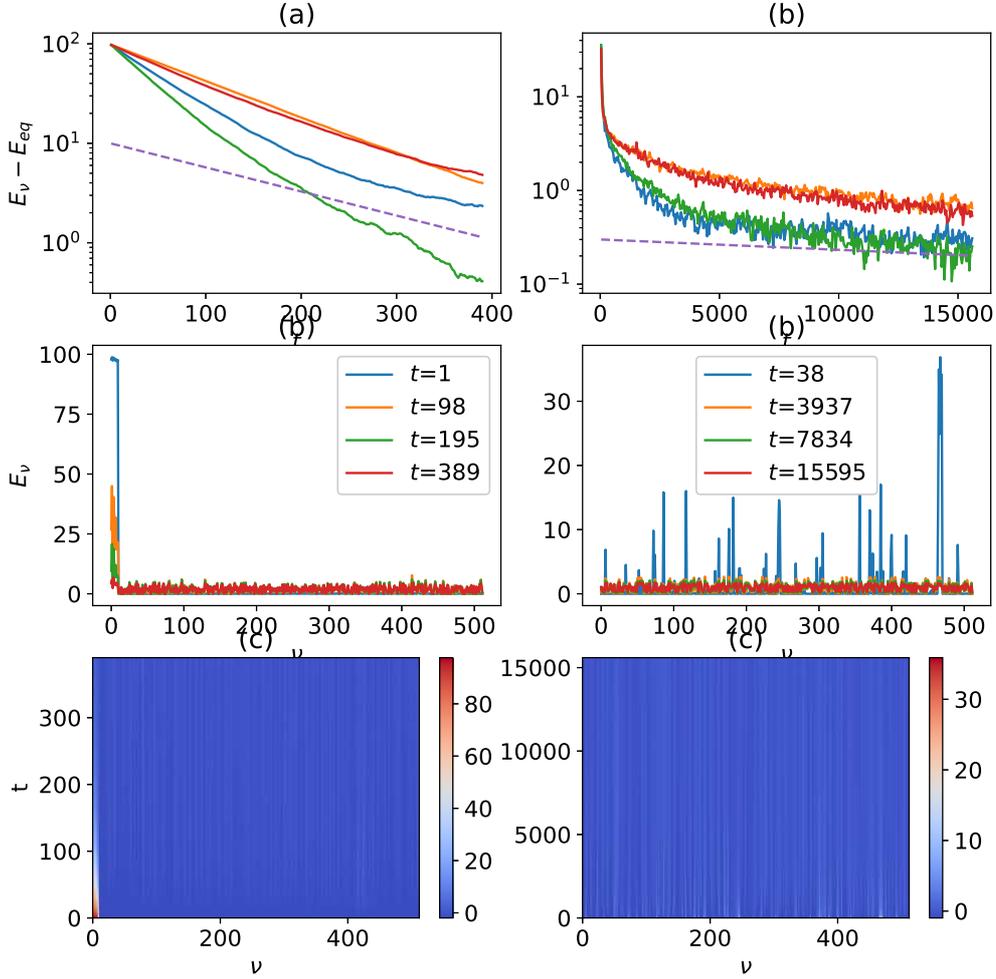} 
\end{center}
\caption{Disordered chain (\ref{dishami}), for $w=2$ (left panels)
$w=10$ (right panels) $\gamma=0.1$ with nearest-neighbor collisions.
initial wavepacket lowest-frequency $N_0=5$ modes
starting from $\nu=0$ and $465$ respectively. 
Upper panels (a,d): solid lines are $E_\nu(t)-E_{eq}$,  
dashed line corresponds to $0.3\exp(-|\mu_2| t)$.
Middle (b,f): snapshots of the mode distributions at different times 
during the thermalization process.
Lower panels (c,g):  evolution of distributions $E_\nu(t)$ in mode space, 
\textcolor{black}{equipartition corresponds to uniform color (up to fluctuations)}.
Data are averaged over 2000 consecutive collisions and an ensemble 
of about 100 initial conditions, with different choices of $\alpha_\nu$
in (\ref{Aini}).  Notice the difference in time scales between left and right panels.}
\label{fig:dterm}
\end{figure}

In figure \ref{fig:dterm} we report the simulation data of (\ref{dishami}) 
for nearest-neighbor collisions and  two values of the disorder parameter $w$,
starting with all the energy fed into a few modes. 
In the upper panels we show the approach to equipartition of some 
of the excited modes.  After an initial transient the decay sets to the 
exponential behavior with a rate that matches fairly well the 
value of the eigenvalue $\mu_2$ (dashed lines). 
The mode evolution and repartition is illustrated in the lower panels.
For weaker disorder, there is a steady transfer from the initial modes to
the others that leads to a uniform depletion of their energies. On the other 
hand, for stronger disorder the transfer is in a sense, more irregular and 
involves modes with  $\nu$ values  different from $\nu_0$.
This is qualitative agreement with what we would expect in view of 
the different structures of the collision matrices as given in 
figure \ref{fig:lyapdis} and \ref{fig:anet} and the different
connectivity in action space.

\begin{figure}
\begin{center}
\includegraphics[scale=0.5]{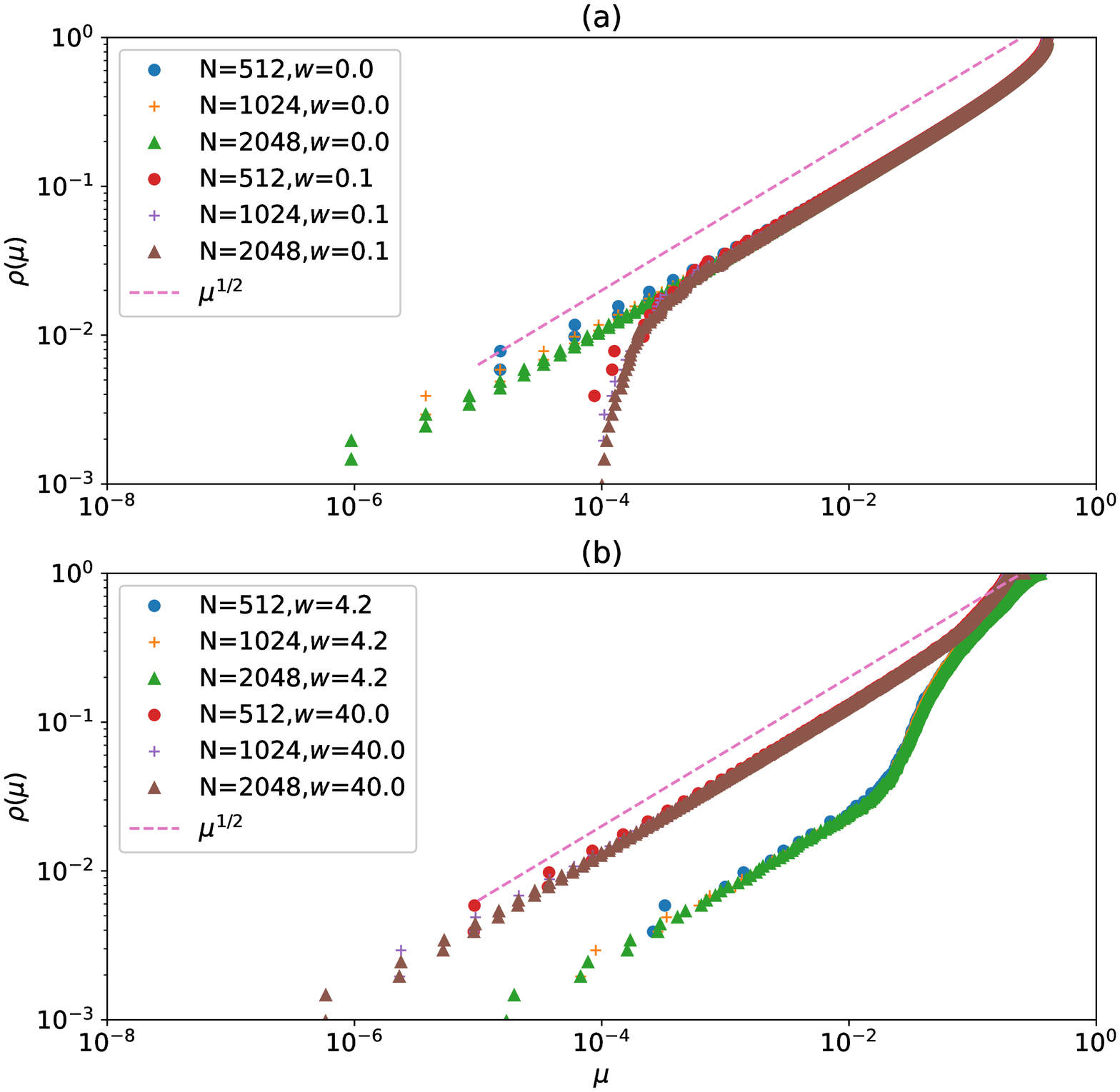} 
\end{center}
\caption{The integrated spectral densities $\rho(\mu)$ 
\textcolor{black}{(i.e. the fraction of eigenvalues less than $\mu$)}
of the collision operator for model (\ref{dishami}).
Plots are for different disorder strengths and three
lattice sizes $N=512,1024,2048$; $\gamma=0.1$
case with nearest-neighbor collisions.
(a) Weak disorder: Comparison between with $w=0$ (ordered case) and 
$w=0.1$ ,  and sizes $N=512,1024,2048$;
(b) Intermediate and strong disorder, $w=4.2, 40$.
The dashed line correspond to the 
scaling  $\sqrt{|\mu|}$ expected for the ordered lattice.}
\label{fig:idos}
\end{figure}

Another observation from figure \ref{fig:dterm} is that the time scales 
involved are much smaller in the case of weaker disorder. 
To understand this issue,  
In figure \ref{fig:idos}a we compare the integrated spectral density
$\rho(\mu)$ of  (\ref{dishami}) in the case with no disorder $w=0$  and 
$w=0.1$,  and for different sizes. As expected, in the ordered case, 
$\rho\sim \sqrt{|\mu|}$, i.e $d_\mu =1$. 
Remarkably,  introducing a weak randomness $\rho$ does
approach zero for a finite $\mu$., i.e the spectral gap is finite, 
meaning that $d_\mu=\infty$ and that the  
relaxation rate is finite and $N$-independent. 
This result should be however be taken with some 
care in the light
of \ref{fig:idos}b that reports the integrated spectral density
for stronger disorder. In the first case, the crossover to 
the  $\sqrt{\mu}$ occurs at very small values of $\mu$.
It is thus possible that such a crossover will occur to 
smaller and smaller $\mu$ values upon decreasing the 
disorder, making $d_\mu$ converge to one in the thermodynamic
limit.

\begin{figure}
\begin{center}
\includegraphics[scale=0.7]{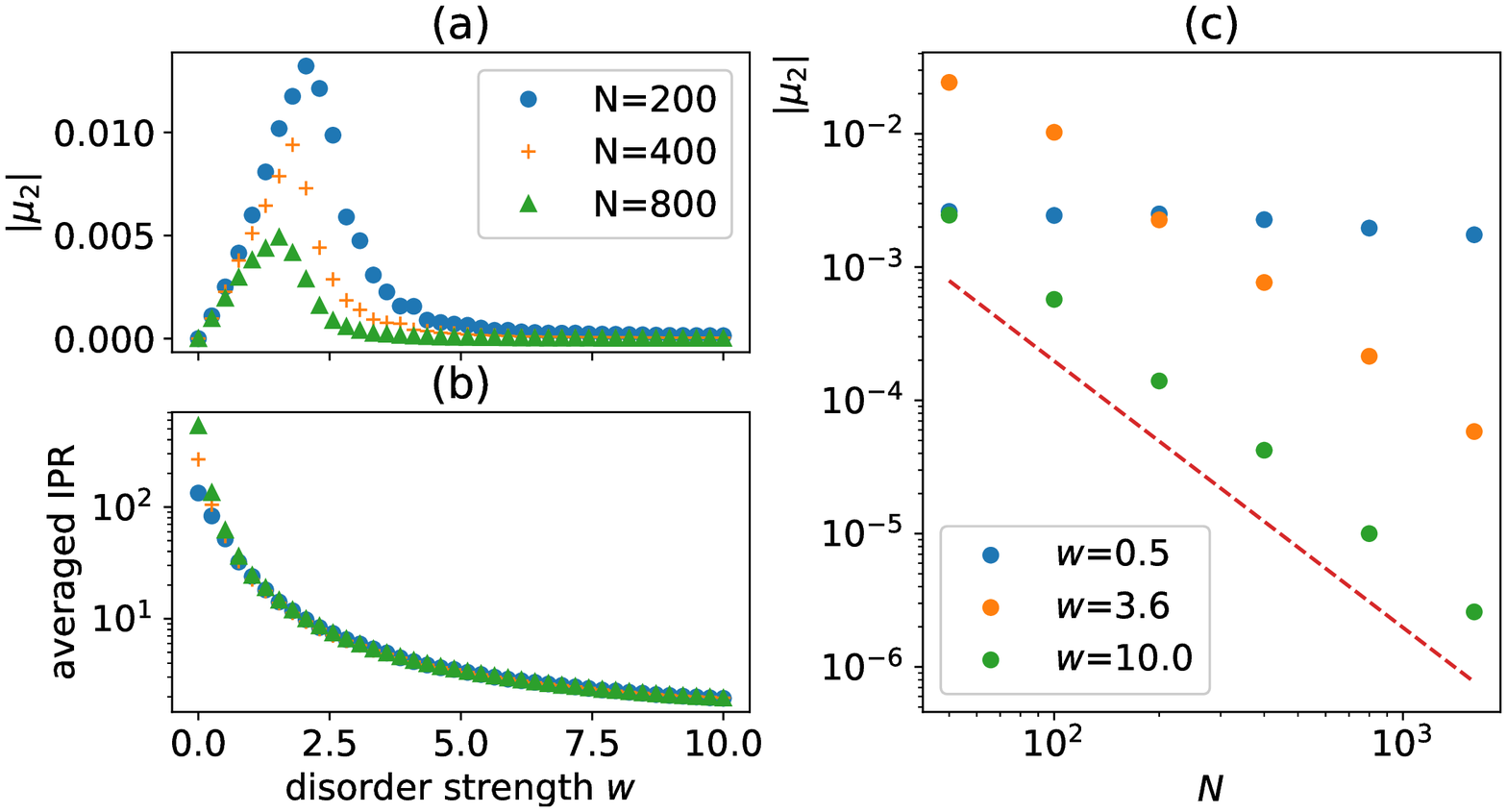} 
\end{center}
\caption{Disordered chain (\ref{dishami}), 
$\gamma=0.1$ with nearest-neighbor collisions.
(a) The absolute value of the second eigenvalue $\mu_2$ 
of the  averaged collision matrix
(\ref{avmat}) and (b) the average inverse participation ratio 
(\ref{ipr}) as a function of disorder strength for 
different chain lengths; (c) reports the 
dependence of the eigenvalue on the lattice size $N$
for different $w$. The dashed line correspond to the 
smallest non-zero eigenvalue of (\ref{diffusion}), 
$\gamma (2\pi/N)^2/2$.}
\label{fig:mu2}
\end{figure}

To further characterize the effect of increasing disorder,
in figure \ref{fig:mu2}a we report the 
spectral gap for different lattice lengths as a function of $w$
along with the average inverse  participation ratio (figure \ref{fig:mu2}b), 
defined as 
\be
IPR=\left[\frac{1}{N}\sum_{\nu} \sum_n \left|\chi_n^\nu\right|^4
\right]^{-1}
\label{ipr}
\ee
which is $O(1)$ for localized modes and $O(N)$ for extended ones.
It thus gives a rough measure of the average localization length.

From the numerical data, it is seen that there exist a value $w_*(N)$ 
such that for $w<w_*$ the spectral
gap is almost  $N$-independent. This confirms that the relaxation 
is faster for weak enough disorder.  On the other hand,  for $w>w_*$ 
the spectral gap vanishes as $N^{-2}$, figure \ref{fig:mu2}c.
A possible argument to estimate the crossover value $w_*$ is to identify
it with the point at which the localization length becomes of the order 
the lattice size.  Above such value some of the coefficients $V_\nu$ start to
become exponentially small and the collision matrix more and more sparse 
accordingly.  This estimate roughly correspond to the location of the 
maxima of $|\mu_2|$ in figure \ref{fig:mu2}a. Indeed, 
using the average IPR as a measure of the localization length, 
we see that $w_*$ should decrease with  $N$,
in qualitative agreement with the leftmost shifting of the 
maxima in figure \ref{fig:mu2}a. 

The $1/N^2$ scaling of the spectral gap suggests that the 
thermalization is ruled by some form of diffusive process in 
action space. This can be understood qualitatively as follows. 
In the  strong disorder case, $w\to \infty$, one can to 
a first approximation,  neglect the coupling in (\ref{dishami}) and consider
the  eigenmodes as localized on a single site.  We are thus in 
close to the case of independent oscillators and the 
matrix $K$ can be approximated, up to exponentially small 
terms, by (\ref{indep}). So, considering the 
case of nearest-neighbor collisions, we argue that,  
up to a re-ordering of 
the eigenmodes indexes $\nu$ by their spatial positions, 
(\ref{releq}) is approximated as
\be
\dot E_\nu=\frac{\gamma}{2}(E_{\nu +1}+E_{\nu -1}-2E_{\nu}),
\label{diffusion}
\ee
which immediately shows that the relaxation is a Brownian
diffusion process in action space. Thus thermalization
starting from a bunch of nearby modes
would entail an initial growth of the number of excited modes 
proportional to $\sqrt{\gamma t}$. For a finite chain 
the longest time scale is given by the smallest 
eigenvalue of   (\ref{diffusion}) which is of order 
$\gamma(2\pi/N)^2/2$. This is consistent with the scaling 
of $\mu_2$ as given in figure \ref{fig:mu2}.

Before passing to the next example, we mention the work \cite{basko2014kinetic} where a 
kinetic theory  in a weakly-disordered nonlinear Schr{\"o}dinger 
chain in the regime of homogeneous chaos. The regime there considered, is the one 
of weak enough interaction 
for the normal modes of the linear problem to remain well resolved but 
strong enough for the dynamics to be chaotic for almost all modes. 
In that case, the kinetic equations are nonlinear, 
leading to nonlinear diffusion \cite{basko2014kinetic}.

\subsection{Mean-field chain}

\begin{figure}
\begin{center}
\includegraphics[scale=0.7]{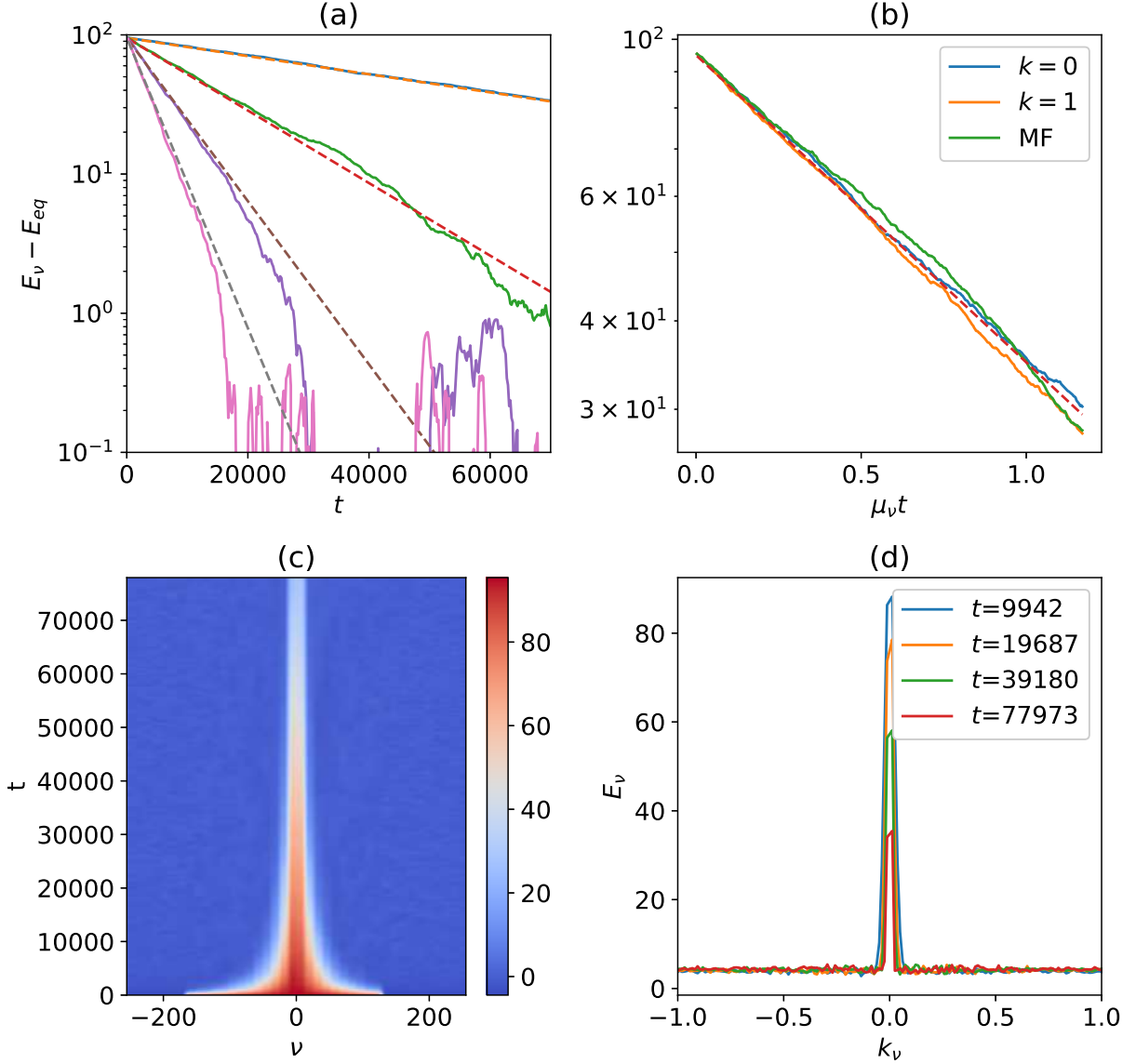}
\end{center}
\caption{Simulations of relaxation to equipartition
for the  chain (\ref{hami}) with nearest-neighbor 
collisions, $k=0$, $N=512$, $\gamma=0.1$:
initial wavepacket consisting of lowest-frequency $N_0=20$ modes. 
(a) Evolution of $E_\nu-E_{eq}$ the four lowest modes 
Dashed lines are  exponential $95\exp(-|\mu_\nu| t)$ with rates 
$\mu_\nu$ given by the approximation (\ref{mu});
plot of dynamics in mode space  (c) and 
snapshots (d) of the mode distribution during the thermalization.
(b)  Evolution of the first mode $E_1-E_{eq}$ as a function 
of $\mu_\nu t$ for 
$k=0$ , $k=1$
and  pure mean-field coupling, dashed line is $95\exp(-|\mu_2| t)$.
Data are averaged over 10000 consecutive collisions and over 
an ensemble of about 100 initial conditions, with different choices of $\alpha_\nu$
in (\ref{Aini}). }
\label{fig:acrel}
\end{figure}

Let us now consider again model (\ref{hami}).
In figure \ref{fig:acrel} we report the data for the 
case for the acoustic chain with $k=0$ and 
nearest-neighbor collisions.
The relaxation of the four lowest modes
is plotted in  figure \ref{fig:acrel}a,b along 
with the snapshots of the evolution of mode energy distributions 
in wavenumber space, \ref{fig:acrel}c,d.
The simulation data are in excellent agreement with 
the relaxation rates computed within the kinetic approach.

From the plots it is seen that thermalization occurs 
by gradual transfer of the energy from the 
excited modes towards all the 
others. Indeed, the energy in the whole background increases
steadily and equipartition is reached with an overall
rate $\mu_2 \sim 1/N^2$ as predicted. 

We also tested the case with collision probability given by
(\ref{wdico}) obtaining similar results. In figure \ref{fig:lyaps}
we plot as red triangles the relaxation rates obtained by
exponential fitting of $E_\nu(t)$: the agreement is 
again very good.

As said, the kinetic equation in this translation-invariant
should be independent of $k$, if the the collision probability 
$W_{n,m}$ is the same. Thus, we simulated the model 
for different $k$ maintaining the collisions only among 
nearest neighbors
In figure \ref{fig:acrel}b we confirm this expectation 
comparing the relaxation of the first Fourier mode for 
three different cases.

\subsection{Elastic network} 

For the Newman-Watts-Strogatz network (\ref{sparse}) we limit 
ourselves to examine the 
spectral gap of the collision operator for different
disorder strengths and different number of sites. 
We just give a first account, leaving a more complete
study of such class of networks to future work.

\begin{figure}
\begin{center}
\includegraphics[scale=0.5]{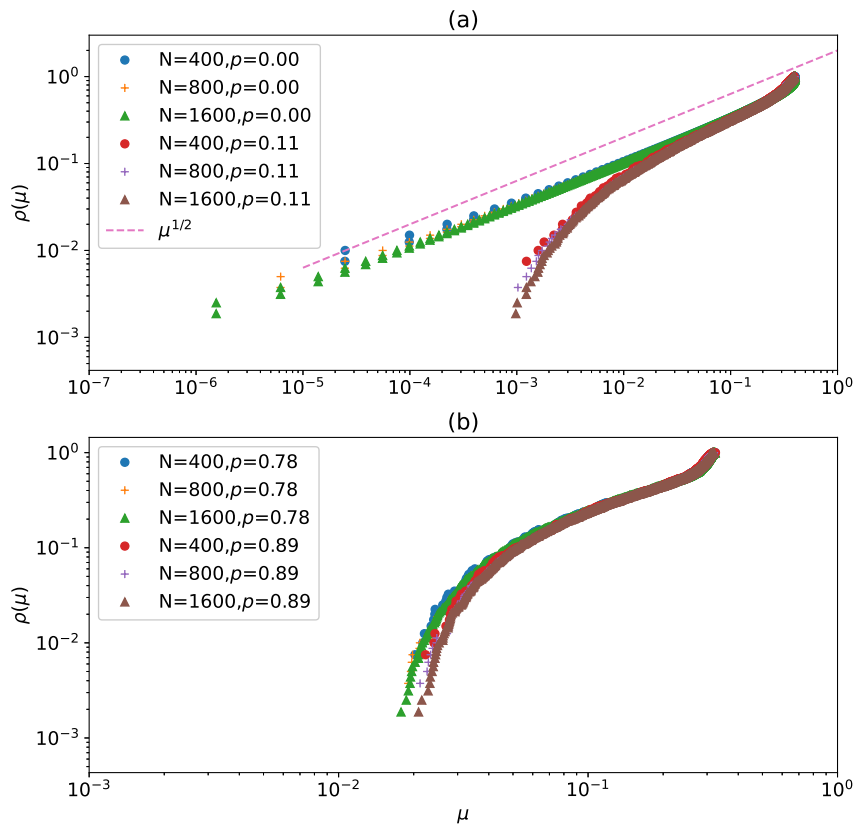} 
\end{center}
\caption{The integrated spectral densities $\rho(\mu)$ of 
the collision operator 
\textcolor{black}{(i.e. the fraction of eigenvalues less than $\mu$)}
for the Newman-Watts-Strogatz
lattices 
for different $p$ and three
lattice sizes $N=512,1024,2048$; $\gamma=0.1$
case with nearest-neighbor collisions.
(a) Comparison between $p=0$ (ordered case) and 
$p=0.11$ ,  and sizes $N=400,800,1600$;
The dashed line correspond to the 
scaling  $\sqrt{|\mu|}$ expected for the ordered lattice.
(b) Intermediate and strong disorder, $p=0.78, 0.89$.
}
\label{fig:idosnws}
\end{figure}

For each $N$ we consider a fixed random realization of
the coupling matrix $C_{i,j}$ and change 
the probability $p$. A few different realizations
of the couplings were examined with qualitatively similar results
indicating that sample-to-sample fluctuations may 
not be very relevant.
In figure \ref{fig:idosnws}a we compare the integrated spectral density
$\rho(\mu)$ of  (\ref{dishami}) 
for different sizes and $p$. 
The data are compatible with a finite spectral gap for $p>0$, 
and accordingly $d_\mu=\infty$, yielding a finite and $N$-independent 
relaxation rate. 

\begin{figure}
\begin{center}
\includegraphics[scale=0.7]{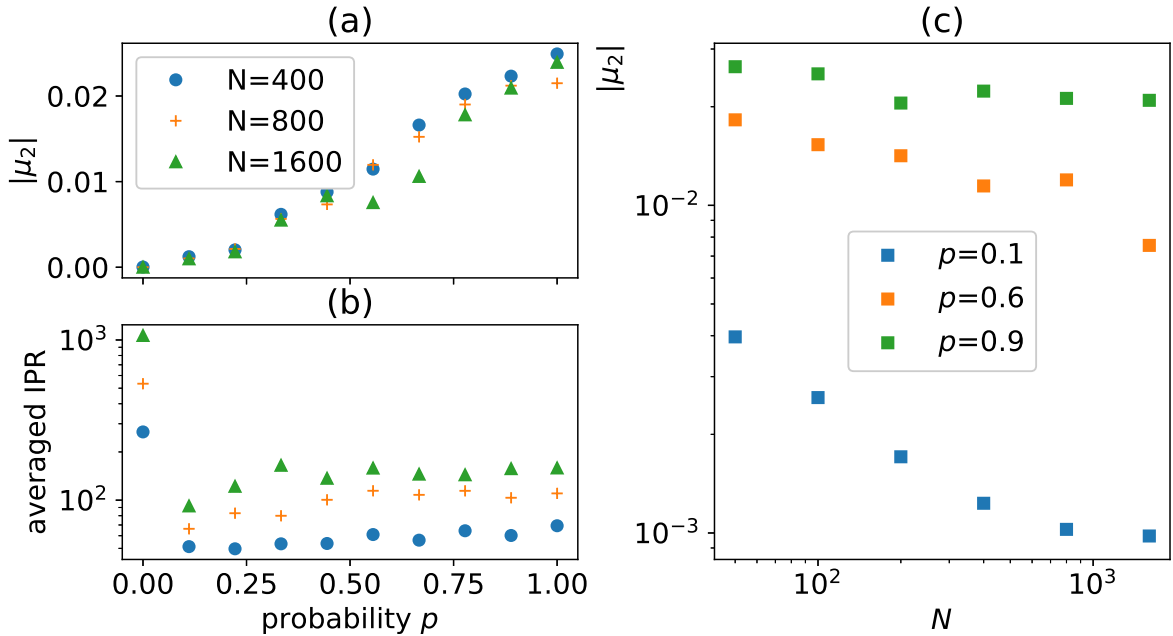} 
\end{center}
\caption{Newman-Watts-Strogatz network (\ref{sparse}), 
$\gamma=0.1$ with nearest-neighbor collisions.
(a) The absolute value of the second eigenvalue $\mu_2$ 
of the  averaged collision matrix
(\ref{avmat}) and (b) the average inverse participation ratio 
(\ref{ipr}) as a function of disorder strength for 
different chain lengths; (c) reports the 
dependence of the eigenvalue on the lattice size $N$
for different $w$.   }
\label{fig:mu2s}
\end{figure}

In figure (\ref{fig:mu2s})a,b
we plot the spectral gap as a function of $p$ 
along with the IPR.
The situation is similar to the disordered chain in the 
weak-disorder regime
see again figure \ref{fig:mu2}. The spectral gap is 
finite for any finite $p$ and closes for $p\to 0$.
We do not observe a crossover to $1/N^2$ or similar 
scaling as in the case of the  disordered chain.

\section{Spectral entropy}

A key indicator that has been often employed 
to characterize the thermalization is the 
non-equilibrium (Shannon) spectral entropy 
\cite{livi1985equipartition}
\be
S(t) = -\sum_\nu E_\nu \log E_\nu; \qquad \sum_\nu E_\nu=1,
\label{entropy}
\ee
where we have fixed the total energy to one, without loosing generality.
For large $t$, $S$ approaches its equipartition value $\log N$. In general
$S$ will depend on the initial conditions. 

\begin{figure}
\begin{center}
\includegraphics[scale=0.6]{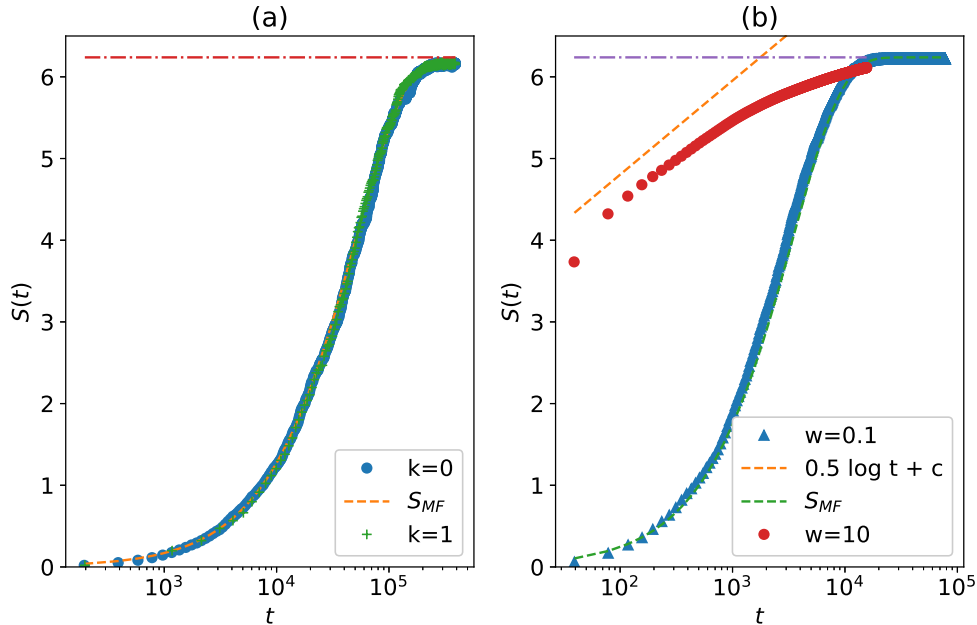} 
\end{center}
\caption{Spectral entropy (\ref{entropy}), 
$\gamma=0.1$, $N=512$ for the harmonic and disordered chains 
with nearest-neighbor collisions.
(a) model (\ref{hami}) with $k=0,1$  and (b) (\ref{dishami}) with 
two values of the disorder. Initial conditions with only 
the lowest frequency mode is excited. Dashed lines correspond
to the approximation (\ref{appentropy}); dot-dashed (red and violet) 
horizontal lines correspond to the equipartition value $S=\log N$.}
\label{fig:sentropy}
\end{figure}

As we have shown above, the energy transfer processes in the translation-
invariant models occurs through a gradual, global, redistribution of the initial energy 
towards all the other modes. In the intermediate times, and assuming 
all the energy initially in one single mode $E_{\nu_0}(0)=\delta_{\nu,{\nu_0}}$ for simplicity,
one could thus perform a kind of "mean-field" approximation  
\cite{mulken2017information}. \textcolor{black}{
By this we mean that the energy of the initially excited mode 
is evenly redistributed among the other $N-1$ ones,} namely $E_\nu\approx (1-E_{\nu_0})/(N-1)$
for $\nu\neq{\nu_0} $
yielding the approximation
\be
S(t) \approx  S_{MF}\equiv -E_{\nu_0} \log E_{\nu_0}-(1-E_{\nu_0})
\log\left(\frac{1-E_{\nu_0}}{N-1}\right)
\label{appentropy}
\ee
where $E_{\nu_0}(t)\approx (1-1/N)\exp(-\mu_{\nu_0} t)+1/N$. 
In figure \ref{fig:sentropy} we plot the simulated 
time evolution of $S$ for the harmonic and disordered chains.
The above formula accounts very well for the data.

The situation is instead pretty different in the case of 
the strongly disordered chain (see the full circles in 
figure \ref{fig:sentropy}b). Here, the action network is different
yielding a diffusive  process. Thus the entropy grows 
logarithmically in the intermediate time range. 
This is readily understood since, as said above, the number
of excited modes grows as a $\sqrt{t}$.  
This fits with the  general fact, that there exist
a close relation between the growth of $S$ and the spectrum of the 
master equation \textcolor{black}{for diffusion 
on networks. In this context, it is known that 
if the spectral density of the Laplacian operator of the graph 
shows scaling with a finite spectral dimension $d_s$, 
the Shannon entropy grows as $\frac{d_s}{2}\log t$ 
with time before reaching the steady-state (uniform) value 
}
(see e.g. \cite{mulken2017information} and the references therein) . 

\section{Discussion and perspectives}

The conservative-noise dynamics for binary collision 
is a succession of random reflections in action space, see equations (\ref{DP})
and (\ref{Aprime}). 
Averaging over random phases yield  
a kinetic equations for the actions/energies of normal modes.  
The mathematical advantage of such dynamics is that the collision operator is linear
and yields a linear master equation (\ref{releq}).  
But the main merit, for our purposes, is that it was possible to construct
explicitly the action network in term of the eigenvectors and the collision
rule.  

Being a stochastic model there is no doubt that equipartition will be finally
obtained and that ergodicity is insured. However, one has the possibility to study
relaxation of an arbitrary finite network by simply computing the 
spectral gap and its dependence on $N$.  This amounts primarily to
understand how the spectral density of the action network depends
on the underlying connections. \textcolor{black}{We also note 
that momentum conservation does not appear to play a major role 
for thermalization. For instance, both the case of strongly 
disordered case and the translation-invariant lattices display the 
same $1/N^2$ scaling of relaxation rates, despite the fact that
momentum is conserved only in the second case.}

Some comments are in order to conclude.

\begin{itemize}

\item The spectral density $\rho(\mu)$ of the collision operator for 
small $\mu$ is
the main ingredient determining the the equipartition time in a finite 
network, see (\ref{ds}).  A natural question is whether there exist a relation between
$d_\mu$ and the spectral dimension in the case in which $\Phi$  of the 
harmonic network in (\ref{hami}) is Laplacian matrix.
In general, we surmise that the two may not be necessarily related.
This is understood in the case of the translation-invariant chain.
There, $d_\mu$ is controlled
by the collision rule only, as seen from (\ref{diag2}) and can, in principle,
be changed independently of the spectral dimension the Laplacian.
However, we have seen that for the disordered chain that the relation is 
less straightforward and this issue may deserve a further analysis.

\item To what extent are the present results applicable to describe a 
genuine nonlinear network? One may guess that the  
stochastic dynamics it is an idealization 
of a regime where chaos is well developed (i.e. the maximal Lyapunov 
exponent is very large) and homogeneous in space. This  
would occur for very large energy densities and/or strongly
nonlinear networks.  If so, correlations should decay rapidly and 
replacing local chaos by random collisions  would be a reasonable 
assumption.  It is also plausible that the overall structure 
of the action network may be described in the same language 
as here,  although with higher-order connectivities (i.e. 
on a hyper-graph). This
is a problem that deserves further studies.

\item In the translation-invariant case 
the kinetic equation is basically the spatially-homogeneous version 
of the well-known phonon Boltzmann equation
(or Boltzmann-Peierls equation) or wave-kinetic equation
for phonon distribution \cite{spohn2006phonon,Onorato2015}.
The main difference is that the random dynamics 
does not conserve the total number (i.e $\sum_\nu I_\nu$ in our 
notation).  As it is known, a more general 
derivation yields an advection term, proportional 
to the phonon group velocity \cite{spohn2006phonon}. 
For the harmonic chain with conserving noise, it has 
been derived in \cite{basile2010energy} with a more 
rigorous approach based on the Wigner distribution.
The difference in our case is that we consider the mode energies and, 
more importantly, we dealt from the very start with the 
homogeneous case, whereby the energy 
distribution is initially spatially uniform and it remains so
a later times. 
The inclusion of such terms is crucial to derive 
the correct hydrodynamic scaling, leading to correct 
(fractional) heat equation \cite{dhar2019anomalous} on longer time-scales (see \cite{basile2016thermal} 
and references therein for 
details).

\item Also, in the language of phonon kinetic theory, a nonlinear potential of 
order $x^p$ induces interaction involving $p$ phonons. 
To the extent in which the random collisions here
described represent an hard-core potential, namely 
an "infinite-order" 
$p\to \infty$ potential, it is thus understandable why 
the phonon modes are globally coupled.  An important difference 
is also that, 
in the standard framework,  energy and momentum conservation imposes 
constraints on the allowed phonon processes:  
(for example conditions like $\omega_w=\omega_k\pm\omega_q$ 
for three-phonons collisions etc. ).

\item In this work we focused on relaxation to equipartition
of the isolated network.
As it is well-known there is a close relation 
between relaxation close to equilibrium  and transport. The relaxation rates 
can be used to predict the correlation decay, and thus
the transport coefficients.  The existence of a 
finite spectral gap of the collision operator is an
indication of fast correlation decay which, in turn,
imply normal diffusive transport. 
In the model we discussed 
this is relatively straightforward once the relaxation
matrix is known. It remains to be investigated how 
its spectral properties depend on the underlying 
connections of the action network. 

\item One may wonder if the same approach can be 
applied to other systems with more conserved quantities.
A relevant example would be the tight-binding electron
problem, modeled as a discrete Schroedinger equation.
In this case the collision rule should preserve both energy 
and the total wavefunction norm. However, it has been
shown that this requires a nonlinear rule 
\cite{iubini2019coupled}. Thus  the mathematical advantages
following from the linearity are lost.

\end{itemize}

The class of models considered here will allow to extend
the study of non-equilibrium properties to the case 
of open networks interacting with external reservoirs \cite{cuneo2018non}. 
We plan to continue this program in the future. 

\vspace{10pt}
\noindent\textbf{Acknowledgements:} I thank Raffaella Burioni, 
Stefano Iubini, Francesco Piazza and Antonio Politi for useful 
correspondences and discussions during elaboration of this 
work.

\vspace{10pt}
\noindent\textbf{Funding:} This work did not receive any funding.

\vspace{10pt}
\noindent\textbf{Data availability:} The datasets generated during and/or 
analysed during the current study are available from the 
corresponding author on reasonable request.

\subsection*{Appendix A: Translation-invariant model}

Using (\ref{Adefi}) we can proceed with the calculation as before 
except for the fact that the normal coordinates $Q,P$ are complex:
%
\begin{gather}
A_\nu= i(2\omega_\nu)^{1/2}Q_\nu+(2\omega_\nu)^{-1/2}P_\nu\\
A_{-\nu}^*= -i(2\omega_\nu)^{1/2}Q_\nu+(2\omega_\nu)^{-1/2}P_\nu,
\end{gather}
and the inverse formulae now read: 
\[
P=\frac12 (2\Omega)^{1/2}\left(A+\tilde{A^*}\right),
\]
where we introduced the shorthand notation $(\tilde A)_\nu=A_{-\nu}$.
Substituting into (\ref{DP}):
\begin{equation}
A'= A - \Omega^{-1/2}VV^+\Omega^{1/2}(A+\tilde A^*)\equiv(1-M)A-M\tilde A^*,
\end{equation}
where the matrix $M$ is given by (\ref{Mcomplex}).
Using equation (\ref{free}) for the free evolution and noticing that
$\tilde A^{*} = e^{-i\Omega \tau}\tilde A^*(t)$ we obtain (\ref{cmapt}).
%



As stated in the text, the formulation of the equation of 
motion in the normal modes coordinates is very convenient
for the implementation of the numerical solution.  
Taking into account the 
form of the  matrix $M$ and that $V_\nu^* = V_{-\nu}$
one can write the collision map as
\be
A' = A- U(W^+A+W^T A^*) 
\ee
where the auxiliary vectors (of size $N$) are 
\[
U_\nu = V_\nu\sqrt{2\over \omega_\nu},\qquad W_\nu = V_\nu \sqrt{2\omega_\nu}
\]
which is convenient for memory allocation as it requires 
only products of one-dimensional arrays at each collision.
Free evolution (\ref{free}) can be equally implemented
as a product if one-dimensional arrays since $\Omega$ is 
diagonal. Therefore this scheme ensures exact energy and 
momentum conservation.

\subsection*{Appendix B: Stochastic equations}

It is useful to rewrite (\ref{cmap}) as a stochastic equation. 
Writing $M\to M dw$ where $dw(t)$ is a stochastic process, which
take the value 0 or 1 at random times, with the Poisson process satisfying
$\langle dw \rangle = dt/\langle \tau \rangle$. 
The infinitesimal changes in the
variables from time $t$ to $t + dt$ to leading order is
\begin{gather}
dA = i\Omega A dt -M\left(A+  A^*\right) dw, 
\label{stoch}
\end{gather}
where $dw^2=dw$ has been used. 
For the translation-invariant case instead 
\begin{gather}
dA = i\Omega A dt -M\left(A+\tilde A^*\right) dw, 
\end{gather}
(see the main text), 
with the same meaning of $dw$.

Equation (\ref{stoch}) can be used to determine the 
equation for the action vector $I=A\circ A^*$, 
where $\circ$ is the Hadamard product (element-wise $A_\nu A^*_\nu=I_\nu$)
with stochastic calculus.
Dropping $dt^2,dtdw$ terms and using again $dw^2=dw$:
\begin{gather}
d(A\circ A^*)= dA\circ A^*+dA^*\circ A + \frac12 dA\circ dA^*\\
=[-M\left(A+  A^*\right)\circ\left(A+  A^*\right)
+M\left(A+  A^*\right)\circ M\left(A+  A^*\right)
]dw, \nonumber
\end{gather}
which contains quadratic, non-diagonal terms.
\textcolor{black}{For completeness, we mention that stochastic
equations have been discussed before for conservative noise, for instance
for weakly interacting anharmonic oscillators \cite{liverani2012toward}.
Here, however, we deal with stochastic dynamics in action space.}

In the kinetic limit, we proceed as in the main text. We 
first perform averaging over uniform random phases
so that only diagonal terms survive $A_\nu A^*_\nu=I_\nu$
in the right-hand side. We get the stochastic equation
for the actions
\be
dI = 2 \left[M\circ M -\mathrm{diag}(M)\right]I \,dw,
\ee
with $\mathrm{diag}(M)$ being the diagonal matrix with diagonal elements $M_{\nu,\nu}$
(again the same symbol $I$ is used for the averages).
Alternatively, we can work with the the mode 
energies defined, in matrix notation, as $E=\Omega I$:
\be
dE = KE\,dw.
\label{dE}
\ee
Equations (\ref{Iprime}) and (\ref{Eprime}) are of course equivalent to 
the above stochastic differential equation formulation.

\bibliographystyle{unsrt}
\bibliography{heat.bib,disorder.bib}

\end{document}